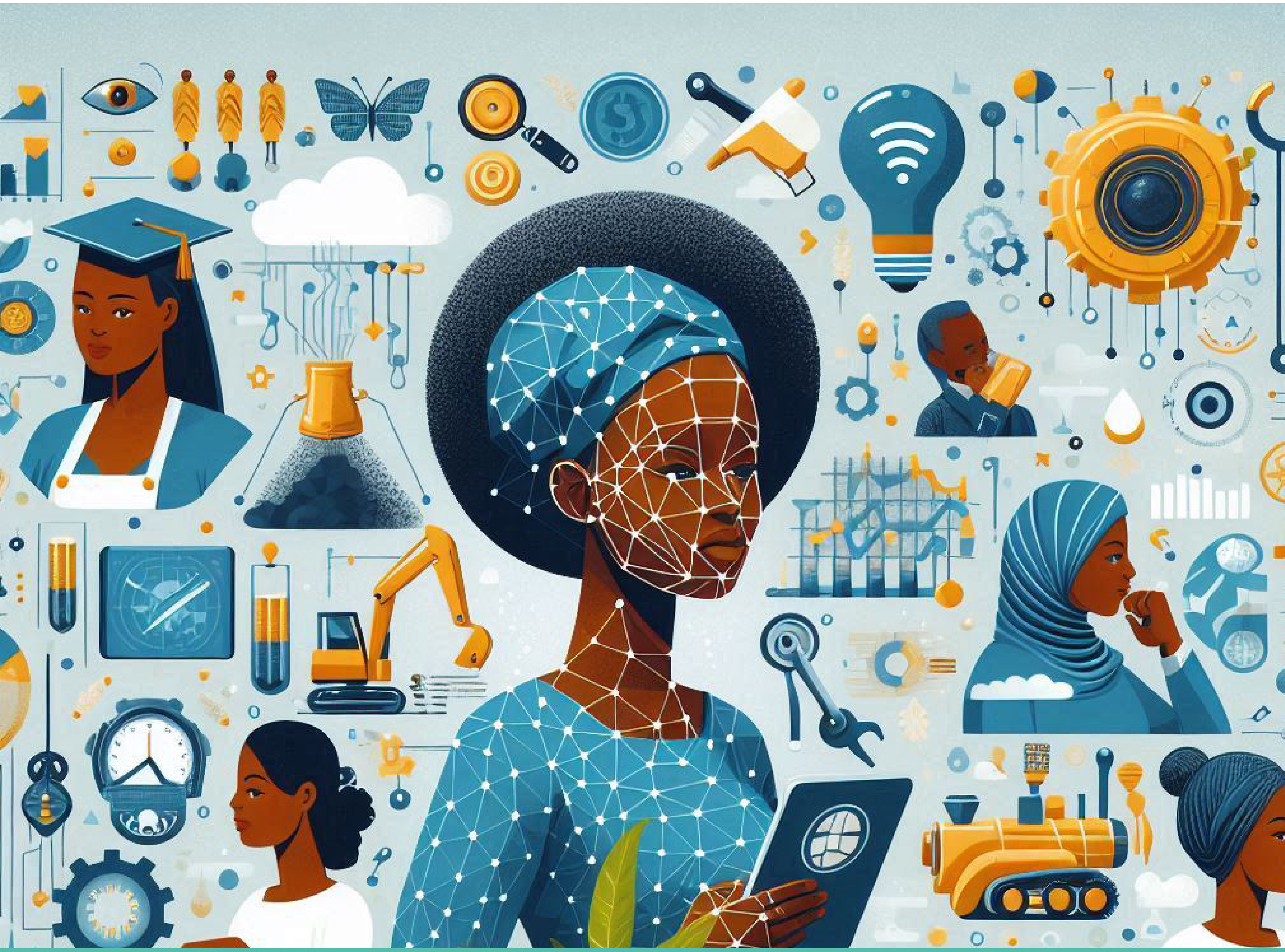

# AI and the Future of Work in Africa
## White Paper

June 2024

# Contents

# Contents

# Executive Summary

This white paper is the output of a multidisciplinary workshop in Nairobi (Nov. 2023). Led by a cross-organisational team including Microsoft Research, Microsoft Philanthropies, University of Pretoria, NEPAD, Lelapa AI, and Oxford University. The workshop brought together diverse thought-leaders from various sectors and backgrounds to discuss the implications of Generative AI for the future of work in Africa. Discussions centered around four key themes: Macroeconomic Impacts; Jobs, Skills and Labour Markets; Workers' Perspectives and Africa-Centric AI Platforms. The white paper provides an overview of the current state and trends of generative AI and its applications in different domains, as well as the challenges and risks associated with its adoption and regulation. It represents a diverse set of perspectives to create a set of insights and recommendations which aim to encourage debate and collaborative action towards creating a dignified future of work for everyone across Africa.

**Brings together a diverse set of perspectives to create a set of insights and recommendations**

## Introduction

The introduction outlines the demographic and socio-economic context in Africa, as it pertains to work. This includes a young population, its often-rural nature and the rich mix of diverse ethnic groups, cultures, religions, and languages. Given this, the African context presents both unique opportunities and challenges, when we consider the potential of Generative AI to positively transform work. This is compounded by the fact that the performance of Generative AI models depends on the amount and quality of training data, yet the majority of the training data for existing generative AI models is sourced from the predominantly English-speaking Global North and as such does not well represent African social and cultural realities.

## Macroeconomic Impacts

The impact of AI on the future that emerges will be a consequence of many things, including technological and policy decisions made today. Getting to a better future will require carefully designed policies and regulations that foster the development of AI while keeping the negative effects in check. This section discusses the potential impacts of Generative AI on three broad areas of macroeconomic interest: productivity growth, labour markets and income inequality, and industrial concentration. It outlines how if leaders wish to maximize the benefits and mitigate the macroeconomic risks related to Generative AI, they must invest in digital infrastructure and human capital - including education initiatives - whilst ensuring that AI development is inclusive and tailored to the continent's unique needs and challenges. Addressing these issues is essential to ensure that AI acts as a catalyst for equitable and sustainable growth in Africa.

## Jobs, Skills and Labour Markets

Africa's young population and vibrant tech ecosystem provide significant opportunities to position Africa as a leader in technological innovation and sustainable development. The section explores the different potential outcomes on labour markets of the deployment of Generative AI – from the potential to enable African youth to forge ahead to potential labour market disruption potentially increasing income inequality. It highlights the need for research which moves beyond the usual generalizations to apply a critical lens to understand the nuances of the repercussions of Generative AI in Africa's unique social and economic contexts. It highlights the importance of:

1. **Preparedness**: Governments, educational bodies, and employers must be agile in reskilling workers. Overarching effort is needed to ensure these transformations improve the quality of the work produced and support and enhance the creativity and value of workers, rather than using AI to automate work – as this will inevitably result in a race to the bottom.

2. **Local AI Leadership**: For Africa to significantly contribute to the AI economy, it is essential to cultivate African talent in AI research, innovation, and design, as well as policy and governance. This of course requires building and consolidating expertise in computer science, machine learning, natural language processing and engineering – the technical skills typically associated with AI development. However, it is clear that such skills are not enough on their own if we are to build AI which enhances human work and creativity. Rather it is important to create environments where multi-disciplinarity can flourish - including the social sciences, ethics, human computer interaction, law and policy – and ensure that diverse perspectives from across society are involved.

3. **Skill Development:** People need the skills, knowledge, and access to leverage Generative AI in their work and careers. Given the tools propensity for fabrication, knowing how to evaluate and appropriately deploy their output will become an important new business skill. Additionally, it is important that the human work of building and maintaining AI systems is recognised and valued as skilled labour.

## Workers' Perspectives

African workers are highly diverse, from urban to rural, frontline to information workers, start-ups to enterprises, and with up to 85% working in the informal sector. The impacts of Generative AI are not likely to be equally distributed across workforces. This section explores what an ideal future might look like for African workers working with Generative AI – including centring African perspectives, work and wellbeing and social contributions. It then goes on to examine the cultural and social alignment and clashes between Generative AI and African perspectives, with a focus on language, context, culture, and data.

- **Language.** Whilst African languages are increasingly represented in large language models (LLMs), they lag behind English performance substantially and currently only a small number are well represented. However, LLMs do perform much better with code-mixed and naturally produced language than previous language technologies, opening up the possibility of better tools in domains such as healthcare and agriculture. However, speech models currently lag behind.

- **Culture and context.** African culture and context are also notably underrepresented in Generative AI training data, leading to poor performance in African workplaces. Representative African data is key to building models which work in African contexts, and this means creating equitable data ecosystems, and incorporating indigenous knowledge in culturally and socially sensitive ways.

- **Data.** The importance of data justice and data sovereignty is highlighted as central to Globally Equitable Generative AI.

Recommendations include centering a communal focus over individualism to balance the needs of communities and individuals; Supporting the substantial informal sector, emphasizing empowerment, entrepreneurship, and job creation over efficiency; Bridging the digital divide, with infrastructure improvements but also edge computing and lower resource AI; Prioritizing sustainable development and well-being; and finding ways to respect and integrate Africa's rich traditional knowledge.

## Africa-Centric AI Platforms

Africa-centric AI refers to the design, development, validation and deployment of AI solutions with a strong focus on African context. The emergence of Africa-centric AI tools and platforms addresses unique socio-economic challenges by tailoring AI solutions to the continent's specific needs. This section discusses a potential dystopian future – where Generative AI exacerbates existing social inequalities - and a potential utopian one – where AI acts as an equalizer. It discusses how the actions taken today determine the future trajectory and how the choice of which world we steer towards is a collective responsibility, requiring engagement from policymakers, technologists, and citizens alike. Ensuring a beneficial outcome with generative AI involves proactive governance, inclusive design, investment in education, and a commitment to regulatory and ethical standards. Recommendations include 1) a commitment to ethical development, transparency, and bias mitigation, 2) Robust regulatory oversight to balance innovation with safeguards against misuse, 3) Encouraging entrepreneurship and innovation, 4) Expanding the grassroots AI communities in Africa, and 5) Learning from communities outside of Africa.

## Summary

Generative AI presents a powerful tool for shaping a dignified future of work in Africa. By proactively addressing the challenges and harnessing the opportunities, Africa can leverage AI to drive economic growth, empower its workforce, and become a leader in socially responsible AI development.

Overall, the recommendations for a Dignified Future of Work for all with Generative AI include

- **Invest in infrastructure and education:** Africa needs strong infrastructure and a skilled workforce to maximize the benefits of AI.

- **Develop inclusive AI policies:** National and regional AI policies focused on inclusive education, worker protection, and stakeholder involvement are essential.

- **Focus on human-centered design:** AI should complement human skills, not replace them. Training data and AI tools should be developed with African contexts in mind.

- **Prioritize African-centric solutions:** Africa-centric AI platforms designed with local expertise can address the continent's specific challenges. Collaboration among stakeholders is key for responsible AI development that respects local knowledge and traditions.

As evidenced in the workshop, *the involvement of youth, community leaders, academics, and business leaders are critical in developing inclusive and relevant AI policies for Africa*. This requires a more agile consultative policy formulation process with sufficient scope for improvement as the Generative AI space evolves. Furthermore, the need for wide disciplinary involvement in the design and building of Generative AI models, platforms and applications is central.

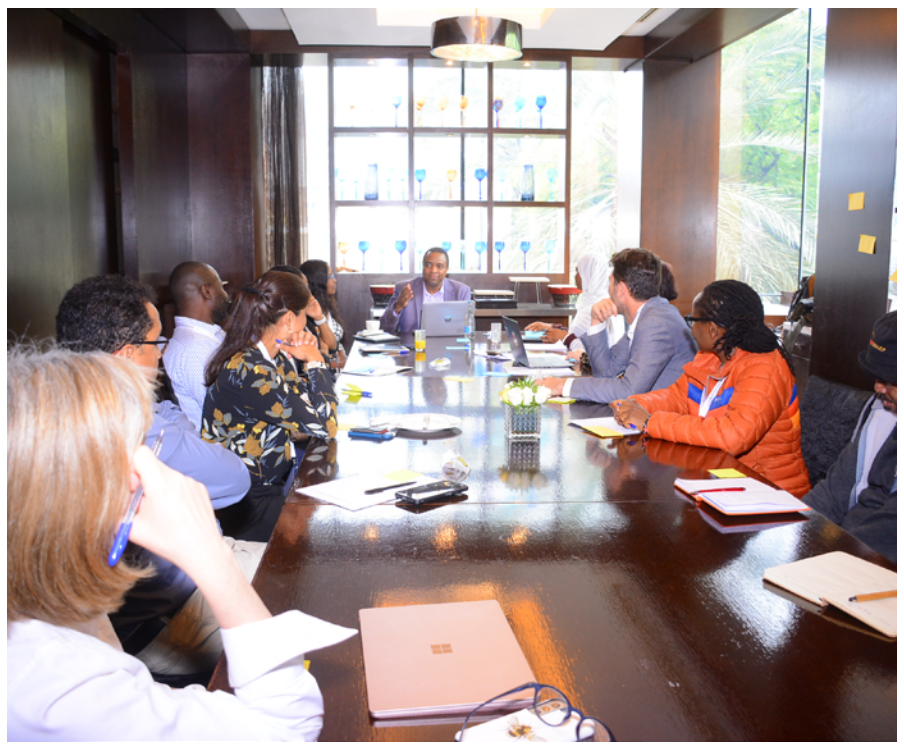
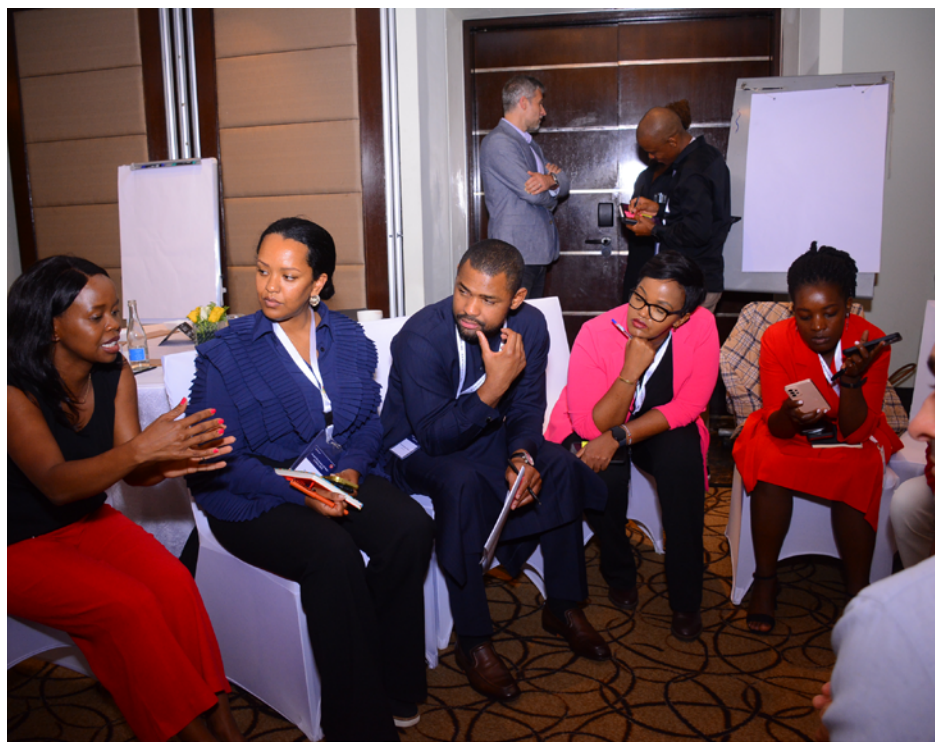
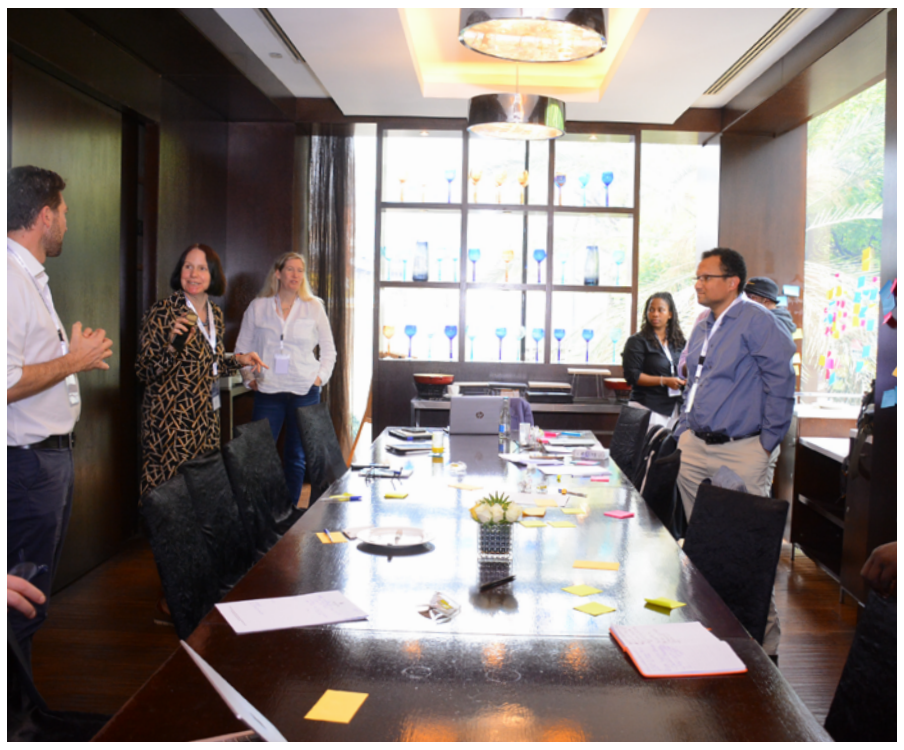
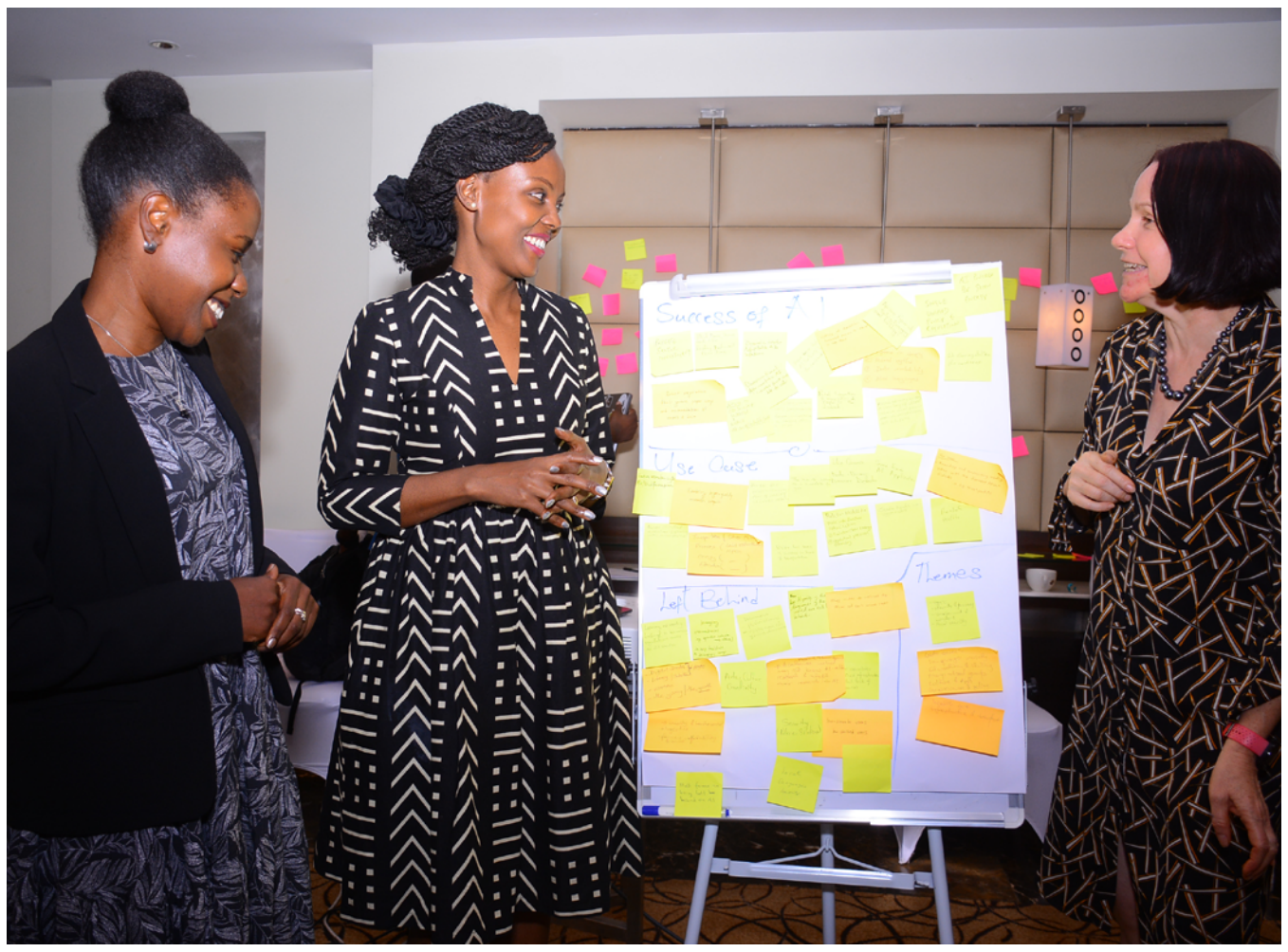
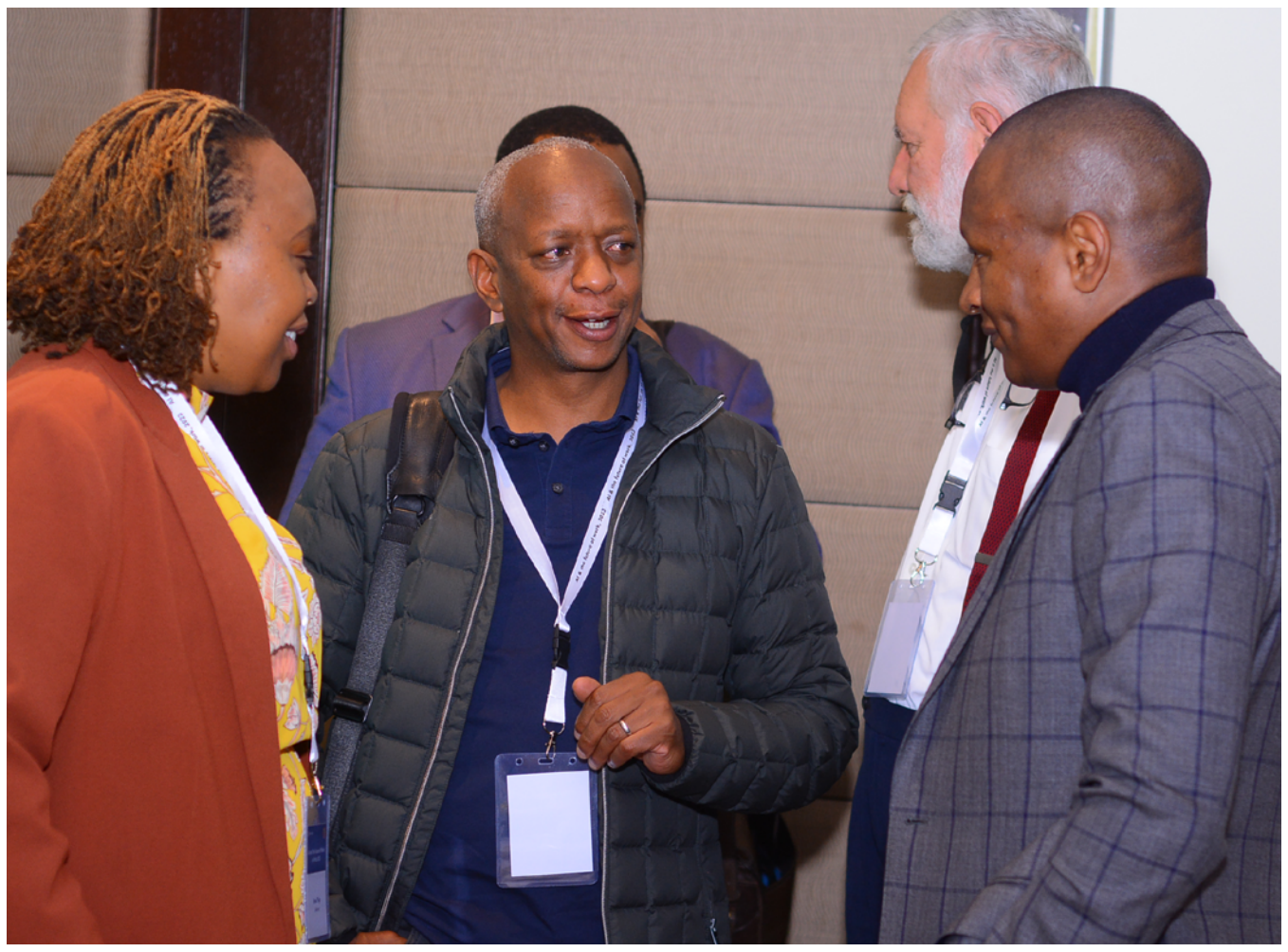
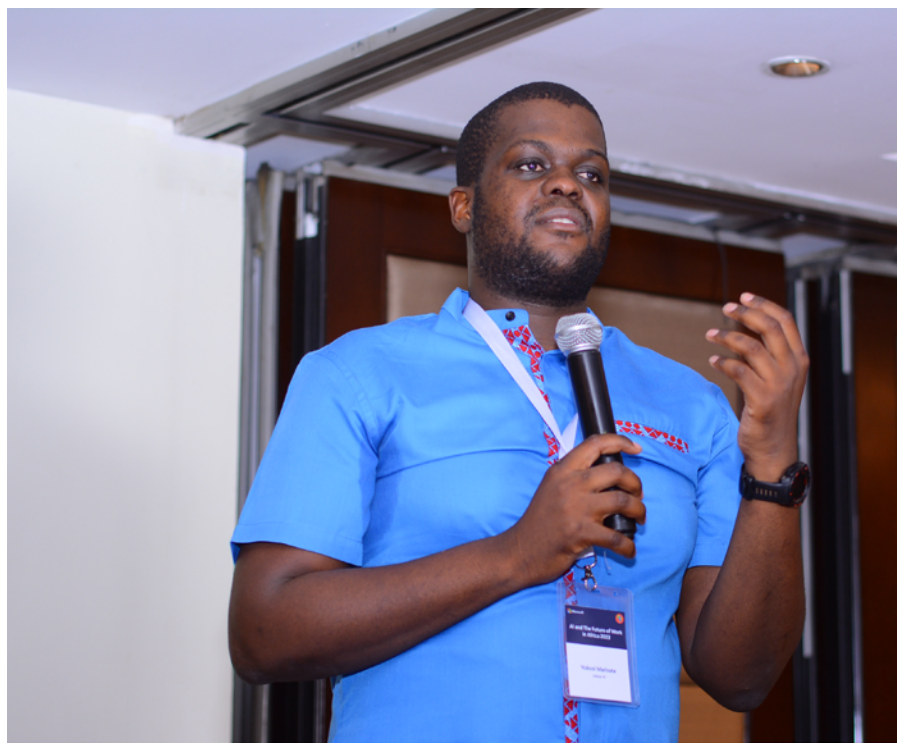
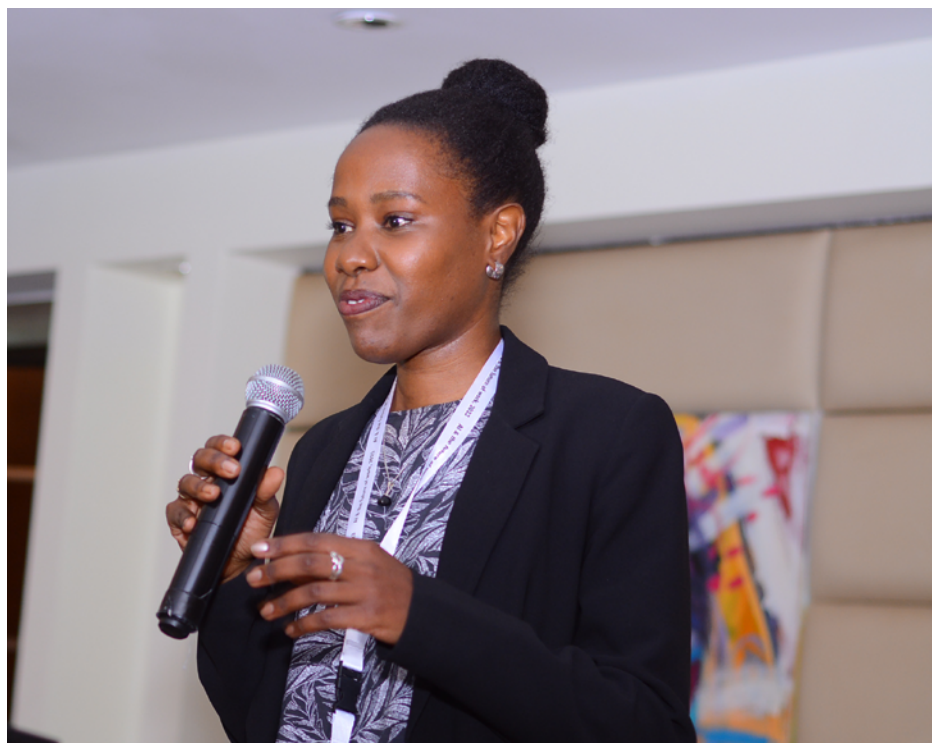
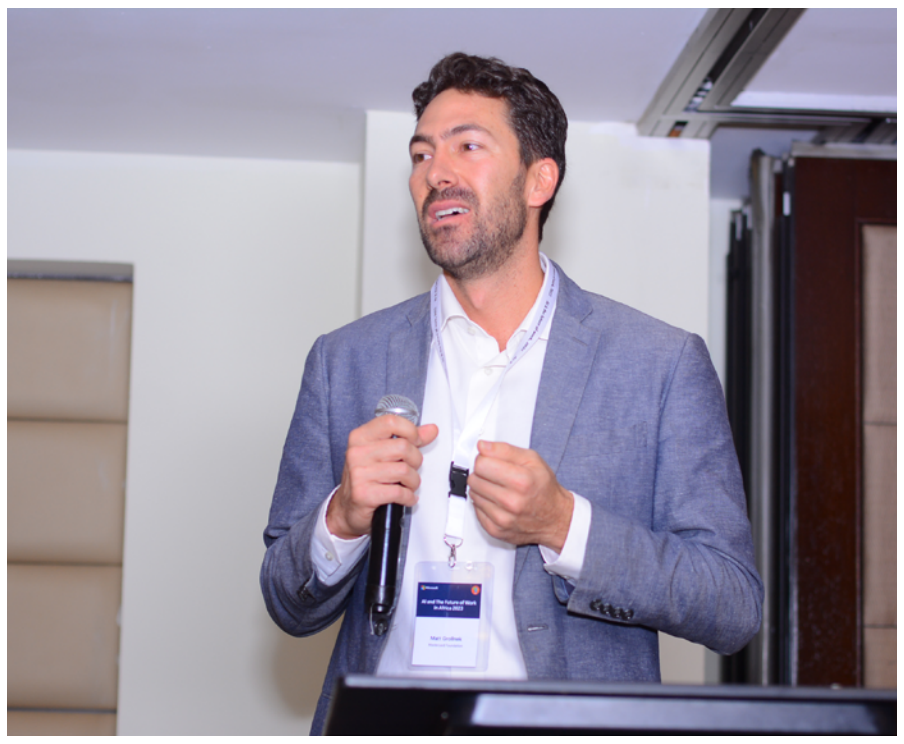
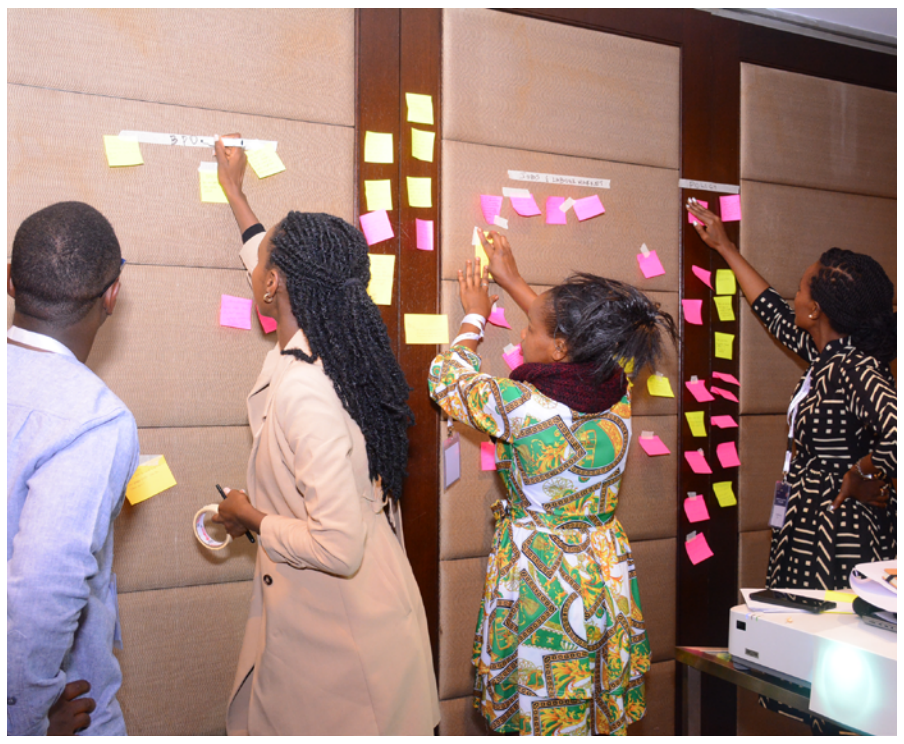
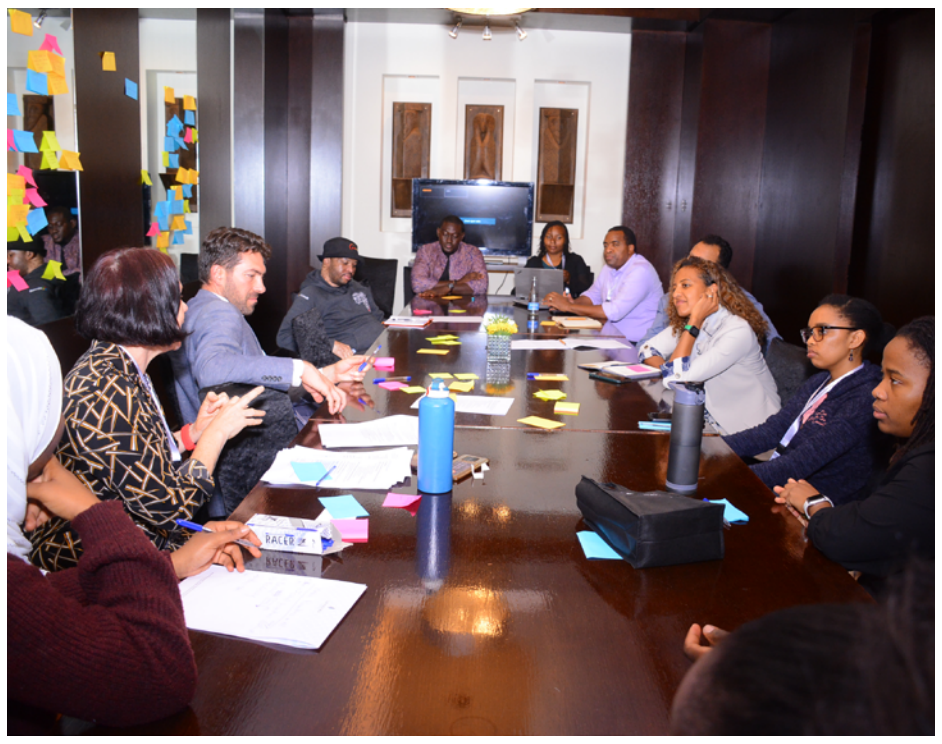
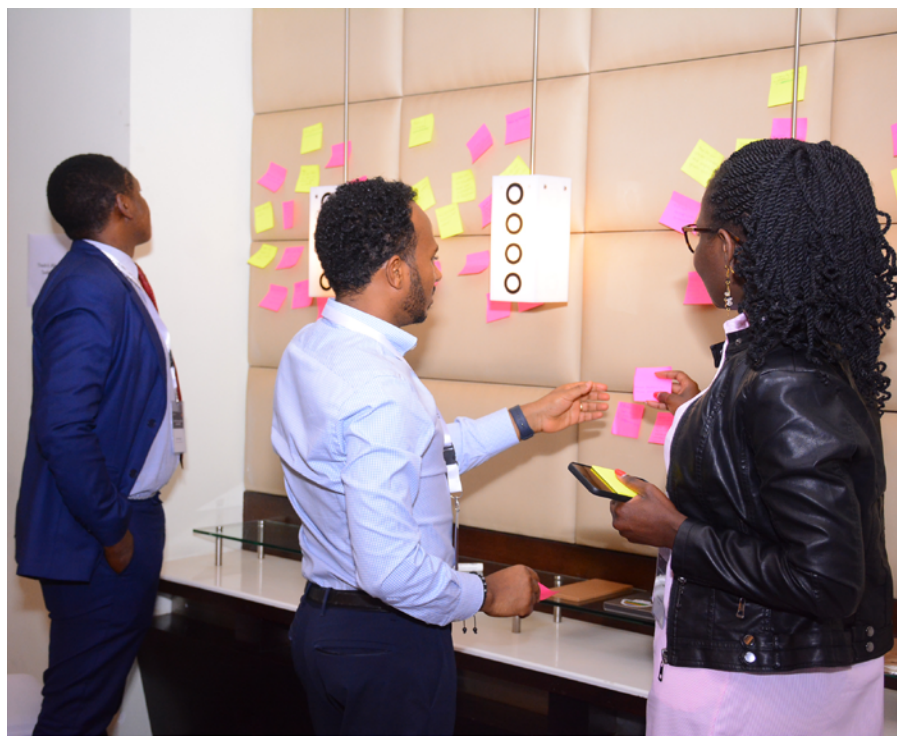

# Introduction

## Pre-Word: About This White Paper

This white paper is the result of a multidisciplinary workshop that took place in Nairobi on 3rd November 2023, where diverse thought-leaders from various sectors and backgrounds discussed the implications of generative AI for the future of work in Africa. The workshop was organized by a core committee including Jacki O'Neill (Microsoft Research Africa, Nairobi), Winnie Karanu (Microsoft Philanthropies), Vukosi Marivate (University of Pretoria and Lelapa AI), Wesley Rosslyn Smith (University of Pretoria), Barbara Glover (NEPAD), Charity Wayua, Matthew Grollnek (Mastercard Foundation), and Anne Makena (Oxford University).

The workshop explored four topics: 1) Macroeconomic impacts, 2) Jobs, skills and labour markets, 3) Workers' perspectives on AI, and 4) Africa Centric AI platforms. This white paper presents the insights and recommendations from these four topics, each written by a different group of contributors.

The white paper aims to provoke discussion and action on whether and how Generative AI can shape the future of work in Africa and how Africa can be at the forefront of designing a dignified future of work for all. The sections of the white paper are complementary but also reflect the diversity of perspectives and backgrounds of the authors. We hope that this diversity will spark further conversations and collaborations on this important topic.

## Why This White Paper Now

Generative AI refers to a subset of artificial intelligence that involves systems capable of creating new content, such as images, text, or even entire datasets. Unlike traditional AI, which relies on explicit programming, generative models are trained on large datasets to learn patterns and generate novel outputs. While there have been many discussions about the potential impact of AI on work in Africa, this paper focuses specifically on Generative AI because it promises to make computing and AI more accessible to a wider range of people.

The ability of generative AI to process natural language and generate new content makes it highly usable for a wide audience across various tasks. For example, users can interact with these models conversationally - asking questions, giving commands, and completing tasks. In the future, this capability could lead to a reduction of complexity across applications and devices, by allowing users to create and find content using natural language without needing to open different applications or even know which tool was used to create the content. Large Language Models (LLMs) could reduce the burden of repetitive and non-essential tasks, such as crafting emails, summarizing documents, and supporting report writing, thereby giving people more time to focus on the work they love. Additionally, multimodal interactions involving image, speech, and video processing and generation further enhance the transformational power of these tools. This could bring AI, and computing more generally, to a wider range of users, including mobile-first or mobile-only users - reaching the billions of people who do not work at desks.

**Generative AI may not be equally useful for everyone, and its impact will not necessarily be evenly distributed globally across regions, communities, and demographics**

As a result, generative AI is likely to transform the future of work across the globe in ways as yet unimagined and has sparked excitement about its potential impact on the United Nations' Sustainable Development Goals (SDGs). However, generative AI may not be equally useful for everyone, and its impact will not necessarily be evenly distributed globally across regions, communities, and demographics. There is a risk of compounding existing systemic inequalities, especially given the imbalances in the training data and data-related processes that typically prioritize data from the Global North.

It is challenging to predict how work will evolve in the next two to ten years. One of the most exciting aspects of this time is that we are at the beginning of this transformation. It is rare to have the opportunity to influence the technologies that will shape our world from the start. Generative AI represents a significant advancement for AI, but it is also very young. The true transformation will come from the multitude of applications built on top of these models.

We have a unique opportunity in Africa to influence what the future of work looks like in these early days when things are not fixed. The question is, what kind of future of work do we want? What do we want technology to do and not do? What does it mean to create a dignified and happy future of work, inclusive of everyone, for both the current and future generations?

## Africa and Generative AI

Generative Artificial Intelligence (AI) has sparked a global conversation about its benefits and impact on the future of work. In African contexts, this conversation brings unique perspectives due to the continent's rich cultural diversity and rapidly changing economic landscape.

### Demographic and Socio-economic Context in Africa

Africa is home to about 1.4 billion people, representing 18% of the world's population, and has the youngest median age of any continent at around 19 years old. Unlike the Global North, which has a high median age and a low growth rate, Africa's population is predicted to double by 2050.

More than 50% of the population of Africa lives in rural areas[1,2], with the majority depending on agriculture for food security and income. Agriculture is the largest economic sector in Africa, accounting for 15% of the continent's Growth Domestic Product (GDP) and employing about 60% of the workforce. Africa also has a large and vibrant informal sector, which accounts for about 55% of the continent's GDP and 80% of the labour force. Nine out of ten rural and urban workers hold jobs in the informal sector, contributing to economic growth, innovation, and resilience.

The continent boasts a rich mix of diverse ethnic groups, cultures, religions, and languages, with well over 1,000 languages spoken, including at least 75 with more than one million speakers. Dominant cultural traits in Africa include a strong sense of community, family, and kinship; respect for elders, ancestors, and nature; a rich oral tradition and storytelling; diverse artistic expressions and forms; and resilience and adaptability to changing circumstances.

1 The world counts. https://www.theworldcounts.com/populations/world/world-rural-population
2 Rural population, percent in Africa | TheGlobalEconomy.com. https://www.theglobaleconomy.com/rankings/rural_population_percent/Africa/

Africa's youthful population includes nearly 1 billion people under the age of 35, with a median age of 18.8 years. By the turn of the century, Africa is estimated to be home to almost half of the world's youth population, nearly twice the entire population of Europe. Currently, about 10-12 million young Africans enter the labor market annually, with only 3 million formal sector jobs available. Whilst primary school enrollment has risen, only 30 to 50% of secondary-school-aged children attend school, and 7 to 23% of young people are enrolled in tertiary education. The lowest levels are found in Central and Eastern Africa and the highest levels in Southern and North Africa[3]. A 2020 report from the International Labour Organization (ILO) indicates that over 20% of African youth are not in employment, education, or training.

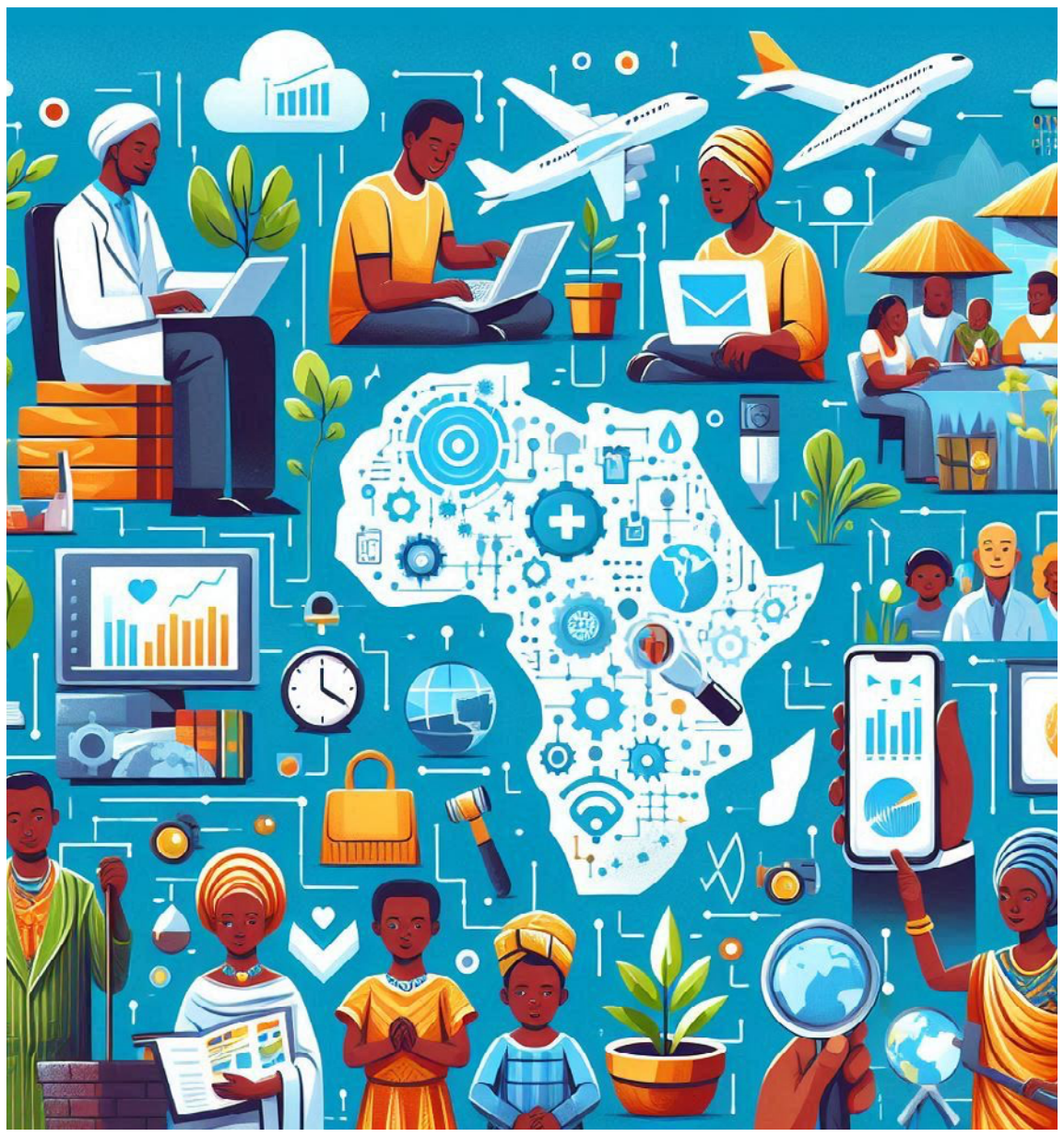

Unemployment rates are concerning in some countries, such as South Africa, where it nears 30%, and the COVID-19 pandemic, coupled with political unrest, has exacerbated the situation in several places.

The African context presents both unique challenges and opportunities for the integration of Generative AI into the future of work. The potential of Generative AI to transform work environments must be considered alongside the continent's demographic and socio-economic realities.

## Generative AI and the African Context

The performance of generative AI models depends on the amount and quality of training data and how decisions about model design and training are taken. When considering performance in African contexts, an important factor is that the majority of training data for existing generative AI models is sourced from the English-speaking Global North. The effectiveness and reliability of AI applications to African work realities are impacted by two key aspects of the training sets:

- **The amount of training data in African languages used by the current crop of Generative AI models[4] is limited**. For example, only Swahili, Afrikaans, Kinyarwanda, and Igbo were declared in the training data for GPT3. This is likely to adversely affect use cases where workers wish to use AI in African languages. It is also likely to disproportionately impact social good use cases where arguably AI might promise the most societal benefit.

- **The amount of training data from African sources used by the current crop of Generative AI models is also limited**, which means that African contexts are likely to be underrepresented in the models[5], impacting performance. Models tend to fail more in situations where use cases are at the tail ends of the training data. This is likely to be a challenge for use in African contexts[6].

3 Musau, Z., 2018. Africa grapples with huge disparities in education. Africa Renewal, 31(3), pp.10-11. https://internationalpolicybrief.org/wp-content/uploads/2023/10/ARTICLE9-115.pdf
4 Including OpenAI models, LLAMA and so on.
5 In part because data from African sources and in African languages is not always readily available for existing training models – for example due to limited availability in open online sources. But also it is important to recognize that even where that data is online it is less likely to have been included and makes a much smaller proportion of the overall training set than data from and about the Global North.
6 See for example, Reflections before the storm: the AI reproduction of biased imagery in global health visuals - The Lancet Global Health or Gondwe, G. 2023. CHATGPT and the Global South: how are journalists in sub-Saharan Africa engaging with generative AI? Online Media and Global Communication. 0 (2023).

## Structure of White Paper

The following sections of the white paper discuss generative AI and the future of work in Africa with a focus on four topics:

1. **Macroeconomics:** This section explores the potential of Generative AI to be a game-changer for African economies, and how its young, growing population create a fertile ground for AI-driven growth. However, it's impact will depend on how it's implemented; its effectiveness hinges on using training data that reflects African languages and contexts; and AI applications must be developed with African realities in mind

2. **Jobs, skills and labour markets:** This section examines how the growing tech scene offer opportunities for AI-driven job creation, but discusses how the continent's lower exposure to AI due to a larger agricultural and informal sector may delay the impact. It calls for a more nuanced examination of the possible impacts of AI on jobs and skills, one which takes into account factors like gender, education, and location.

3. **Workers' perspectives on AI:** Generative AI has the potential to empower African workers, but significant hurdles exist. High data costs and limited access to devices like smartphones and computers restrict internet use and thus, AI tools. However, Africa's strong entrepreneurial spirit positions workers well to leverage AI, as long as development prioritizes African perspectives. Further, AI should complement, not replace, interpersonal communication, a cornerstone of African business culture.

4. **Africa-Centric AI platforms**: This section discusses the importance of Africa-Centric AI systems, presenting bot utopian and dystopian world views, which can be conceived of as arising as a consequence of widespread Generative AI adoption. It examines how the trajectory towards either future is determined by the actions taken today and advocates for proactive governance, inclusive design, investment in education, and a commitment to regulatory and ethical standards.

# Macroeconomic Impacts

## Introduction

Generative AI is expected to have a transformative and lasting effect on economies at a global scale. Africa will not be an exception; however, the integration of Generative AI in Africa presents a unique landscape of both opportunities and risks, impacting macroeconomic growth over the next decade.

Despite stagnant or waning growth across the continent over the last ten years, analysts suggest that Africa has the human capital and resources it needs to propel economic growth and increase productivity across all sectors, particularly the services sector. Africa's population is expected to nearly double to 2.5 billion people by 2050[1] which would make her home to the youngest population on earth. Sound macroeconomic policies can ensure that this demographic dividend is fully realised. Additionally, Africa offers trillion-dollar private sector investment opportunities, including in the climate and green growth sectors[2].

**Africa offers trillion-dollar private sector investment opportunities in the climate and green growth sectors**

The overall impact of Generative AI on macroeconomic development in Africa is yet unknown and multiple scenarios abound, ranging from overly optimistic to exceptionally pessimistic. Experts assess that AI, including Generative AI[3], has the potential to impact three broad areas of macroeconomic interest: **productivity growth, labour markets and income inequality**, and **industrial concentration**[4].

1 Kuyoro, M., Leke, A., White, O., Woetzel, J., Jayaram, K. and Hicks, K., 2023. Reimagining economic growth in Africa: Turning diversity into opportunity. McKinsey Global Institute Special Report, 5.
2 African Economic Outlook. 20241. Driving Africa's Transformation: The Reform of the Global Financial Architecture. African Development Bank Group. Available at: https://www.afdb.org/en/knowledge/publications/african-economic-outlook. [Accessed 7 Jun 2024]
3 Generative AI refers to a subset of artificial intelligence that involves systems capable of creating new content, such as images, text, or even entire datasets. Unlike traditional AI, which relies on explicit programming, generative models are trained on large datasets to learn patterns and generate novel outputs. Notable examples include Generative Adversarial Networks (GANs) and Transformers.
4 The macroeconomics of artificial Intelligence (2023). https://www.imf.org/en/Publications/fandd/issues/2023/12/Macroeconomics-of-artificial-intelligence-Brynjolfsson-Unger.

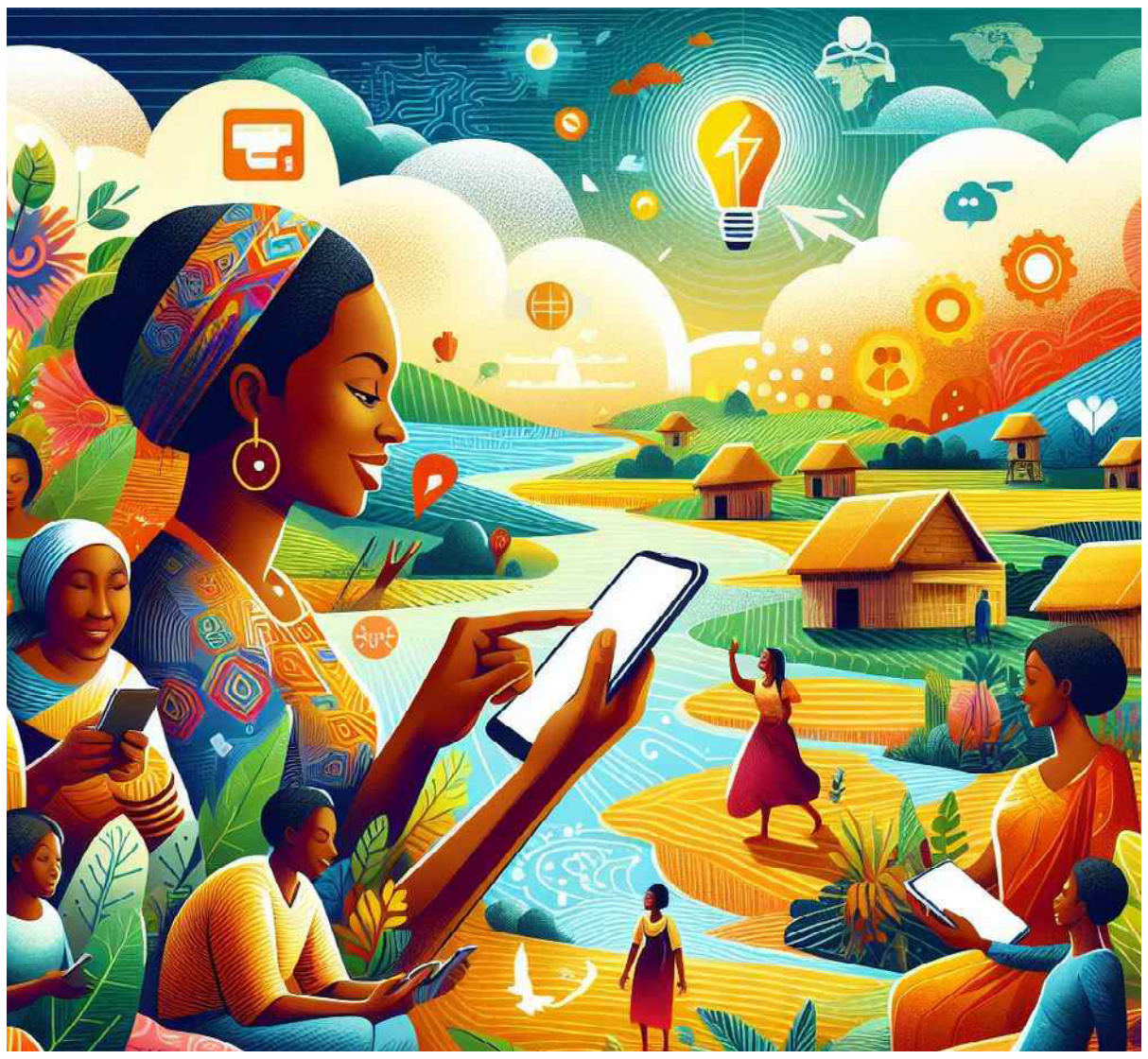

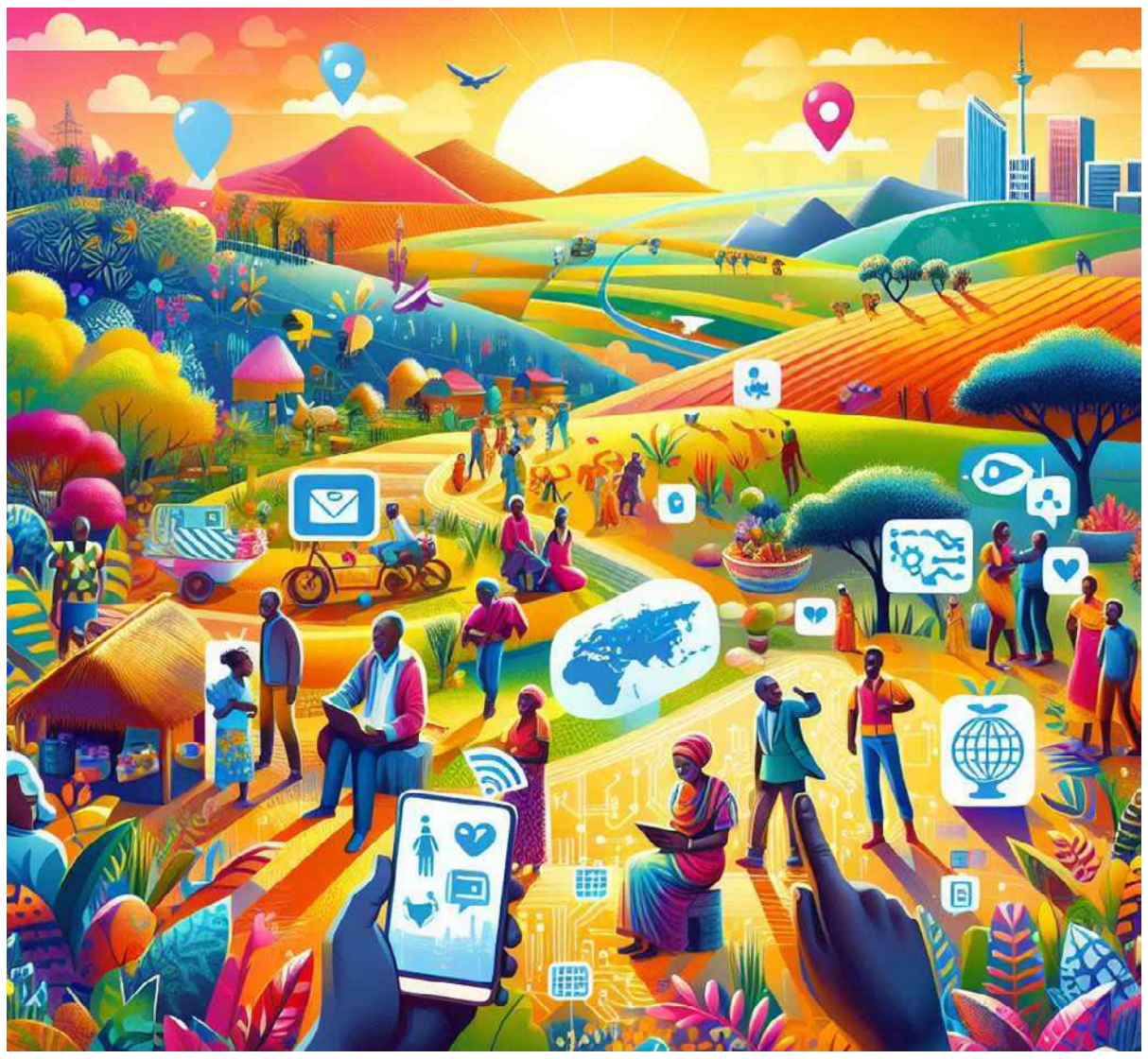

## Productivity Growth

Boosting productivity growth is arguably one of the world's most fundamental economic challenges[5]. But disagreements persist about the extent to which AI will boost productivity, both globally and in Africa. It is estimated that in South Africa alone productivity increase from Generative AI could account for 0.5% of the GDP growth due to automation[6].

AI is most likely to improve productivity in capital-intensive sectors, such as manufacturing and transport, and sectors where routine tasks lend themselves to automation[78].

AI could also help to revolutionize agriculture in Africa, a key sector for most countries across the continent, by accelerating precision farming, improving crop yields and sustainable farming practices, enhancing value and supply chains, expanding access to export markets, and boosting the sector's contribution to GDP. With half of Africa's workforce still employed in farming, boosting agricultural productivity and farmer incomes remains critical[9]. There are around 250 million small-holder farmers across Africa who provide 75% of the food for the continent. Tools such as SMS chatbots to help small-holder farmers with weather, planting, disease and so on have been around for some time[10]. However, with Generative AI's advances in supporting natural language interaction in several African languages the possibility of providing such services at scale becomes more likely in the coming years.

AI could also bring benefits to healthcare and financial services. For example, by improving the delivery of healthcare - through remote care and health data management. If used to extend the reach of services to areas with limited medical personnel, this could prove groundbreaking. Although it will require investment in medical professionals to manage such services. It is important to remember that AI is best used to augment, not replace, human work. AI-powered fintech solutions might enhance financial inclusion, crucial for macroeconomic growth. AI can enable more efficient banking processes, risk assessment for loans, and personalized financial services,

5 The macroeconomics of artificial Intelligence (2023). https://www.imf.org/en/Publications/fandd/issues/2023/12/Macroeconomics-of-artificial-intelligence-Brynjolfsson-Unger.

6 Chui, M., Hazan, E., Roberts, R., Singla, A., Smaje, K., Sukharevsky, A., Yee, L. and Zemmel, R., 2023. The economic potential of generative AI: The next productivity frontier. McKinsey.

7 Szczepanski, M., 2019. Economic impacts of artificial intelligence (AI). https://www.europarl.europa.eu/RegData/etudes/BRIE/2019/637967/EPRS_BRI(2019)637967_EN.pdf

8 World Bank Open Data. https://data.worldbank.org/indicator/NV.IND.MANF.ZS?locations=ZG.

9 Kuyoro, M., Leke, A., White, O., Woetzel, J., Jayaram, K. and Hicks, K., 2023. Reimagining economic growth in Africa: Turning diversity into opportunity. McKinsey Global Institute Special Report, 5. https://www.mckinsey.com/mgi/our-research/reimagining-economic-growth-in-africa-turning-diversity-into-opportunity

10 See, for example, FarmVibes.Bot - Microsoft Research

reaching populations traditionally excluded from formal banking systems. AI might also be leveraged to optimise trade processes, currency valuation, and stabilise regional and national economies, promoting fair trade and sustainable development across the continent.

All kinds of jobs are likely to be impacted by Generative AI – including professions, technical and administrative work and frontline work. For example, some have suggested that clerical jobs, such as a legal assistant, might be eliminated or drastically changed. The education, legal, news and corporate office sectors are also expected to undergo transformations in the type of work done, the skills required, and the outputs produced. McKinsey suggested that Generative AI could enable labor productivity growth of 0.1 to 0.6 percent annually through 2040, depending on the rate of technology adoption and redeployment of worker time into other activities[11].

Despite the promises heralded about AI and productivity, these gains may not be realised if firms fail to make the organisational and managerial changes necessary to best leverage AI for enhanced productivity. The financial costs of deploying Generative AI could be out of reach for many organisations in Africa, where MSMEs who are the economic backbone are least likely to be able to afford such tools. Whilst free versions are available of many tools, MSMEs tend to refrain from using them for business sensitive or confidential data because of lack of clarity about whether that data may be used in future model training.

AI uptake by firms may also be hindered by external factors such as legal regimes that excessively constrain innovation and the use of AI. At the same time, it is vital that policies and practices ensure that AI is deployed responsibly, and that AI related labour is valued and dignified (See Jobs, Skills and Labour Markets Section). Getting the right amount of regulation for the right type of economic growth (where the value is felt across all the different strata of society, rather than concentrated in the hands of a few) will be crucial for the well-being of society.

Even if workers in some low- and middle-income countries (LMICs) have access to individual technologies, the lack of basic infrastructure in many locales - such as affordable and consistent internet and energy access - could limit the potential benefits of technological change for firms and individuals alike and amplify productivity gaps[12].

## Labour Markets and Income Inequality

Despite improvements in the business regulatory environment, informality in Africa has persisted over the last two decades, remaining, on average, at around 75 percent of total employment[13].

Traditionally, lower-paying jobs have been considered most at risk of being replaced by AI and automation, while well-paid skilled jobs were considered likely to be in high demand and safer[14]. However, Generative AI is likely to transform medium and high skilled knowledge workplaces across the board. How this will play out remains to be seen, but it is important to attend to any changes which might lead to higher income inequality. Some studies suggest that Generative AI might primarily displace ‘middle-skill’ workers and lead to growth in the share of low- and high-wage labour, thus widening income inequality[15][16]. Some have suggested that fears of joblessness and unemployment might therefore be misplaced, and greater focus should be directed at the prospect of rising 11 Chui, M., Hazan, E., Roberts, R., Singla, A., Smaje, K., Sukharevsky, A., Yee, L. and Zemmel, R., 2023. The economic potential of generative AI: The next productivity frontier. McKinsey. (p47)

12 Golman, M. and Ernst, C., 2022. Future of Work, Emerging Sectors and the Potential for Transition to Formality.

13 World Bank. 2019. World Development Report 2019: The Changing Nature of Work. Washington, DC: World Bank. doi:10.1596/978-1-4648-1328-3 f

14 Szczepanski, M., 2019. Economic impacts of artificial intelligence (AI). https://www.europarl.europa.eu/RegData/etudes/BRIE/2019/637967/EPRS_BRI(2019)637967_EN.pdf

15 Johnson, S. and Acemoglu, D., 2023. Power and progress: Our thousand-year struggle over technology and prosperity. Hachette UK.

16 Szczepanski, M., 2019. Economic impacts of artificial intelligence (AI). https://www.europarl.europa.eu/RegData/etudes/BRIE/2019/637967/EPRS_BRI(2019)637967_EN.pdf

inequality[17]. There is also rising concern that Generative AI might disrupt the creative sector[18], relegating creative workers to supervising the technology.

If AI helps those with less experience or skills deficits succeed in new jobs or sectors, it could lead to a net increase in employment. And, if AI lives up to the promise of freeing-up time and allowing humans to focus on more creative work, novel solutions, or complex problem-solving, this could result in increased productivity or even better work-life balance. However, it is important to note that technology cannot do this on its own – it is the macroeconomic, labour and regulatory markets in which these technologies operate which will need to be adapted to support positive change.

## Industrial Concentration

What might the growth and expansion of an African AI industry, comprising both foreign and indigenous firms look like?

In one scenario, the industrial concentration of AI across multiple sectors could increase, with only the largest firms (and predominantly those in the Global North) intensively using AI in their core business.

In this scenario, AI might enable these firms to become more productive, profitable, and larger than most competitors in the Global South.

Additionally, if the costs of developing AI models remain as high as they are today—in terms of computational power, data, and human capital—this could increase African countries' dependency on firms outside the continent to provide AI technologies. This could contribute to wider patterns of technological neo-colonialism[19] and AI sweatshops[20].

But, with the appropriate policies and regulations in place— as well as access to the right data, compute and investment— indigenous AI firms could flourish, supporting a wider uptake of AI across the continent, particularly for small and medium-sized enterprises. These firms could use AI to gain deeper market insights, improve supply and value chains, and contribute to macroeconomic growth and labour markets.

**With the appropriate policies and regulations in place indigenous AI firms could flourish supporting a wider uptake of AI across the continent**

17 Jacobs, J. (2023) The macroeconomics of artificial intelligence. https://www.omfif.org/2023/06/the-macroeconomics-of-artificial-intelligence/

18 De Cremer, D., Bianzino, N.M. and Falk, B., 2023. How generative AI could disrupt creative work. Harvard Business Review, 13. https://hbr.org/2023/04/how-generative-ai-could-disrupt-creative-work

19 Tibebu, H. (2024) Why Africa must demand a fair share in AI development and governance. https://www.techpolicy.press/why-africa-must-demand-a-fair-share-in-ai-development-and-governance/.

20 Hellerstein, E. (2024) Silicon Savanna: The workers taking on Africa's digital sweatshops. https://www.codastory.com/authoritarian-tech/kenya-content-moderators/.

# Takeaways and Recommendations

There is no consensus on whether and to what extent the risks related to Generative AI and the future of work will materialise. The potential for AI to support macroeconomic growth in Africa could be significant, particularly in sectors like agriculture, healthcare, services, and manufacturing. However, this potential comes with significant risks, including labour displacement and greater industrial concentration with an increasing dependence on technology firms in the Global North. Neither outcome is a given.

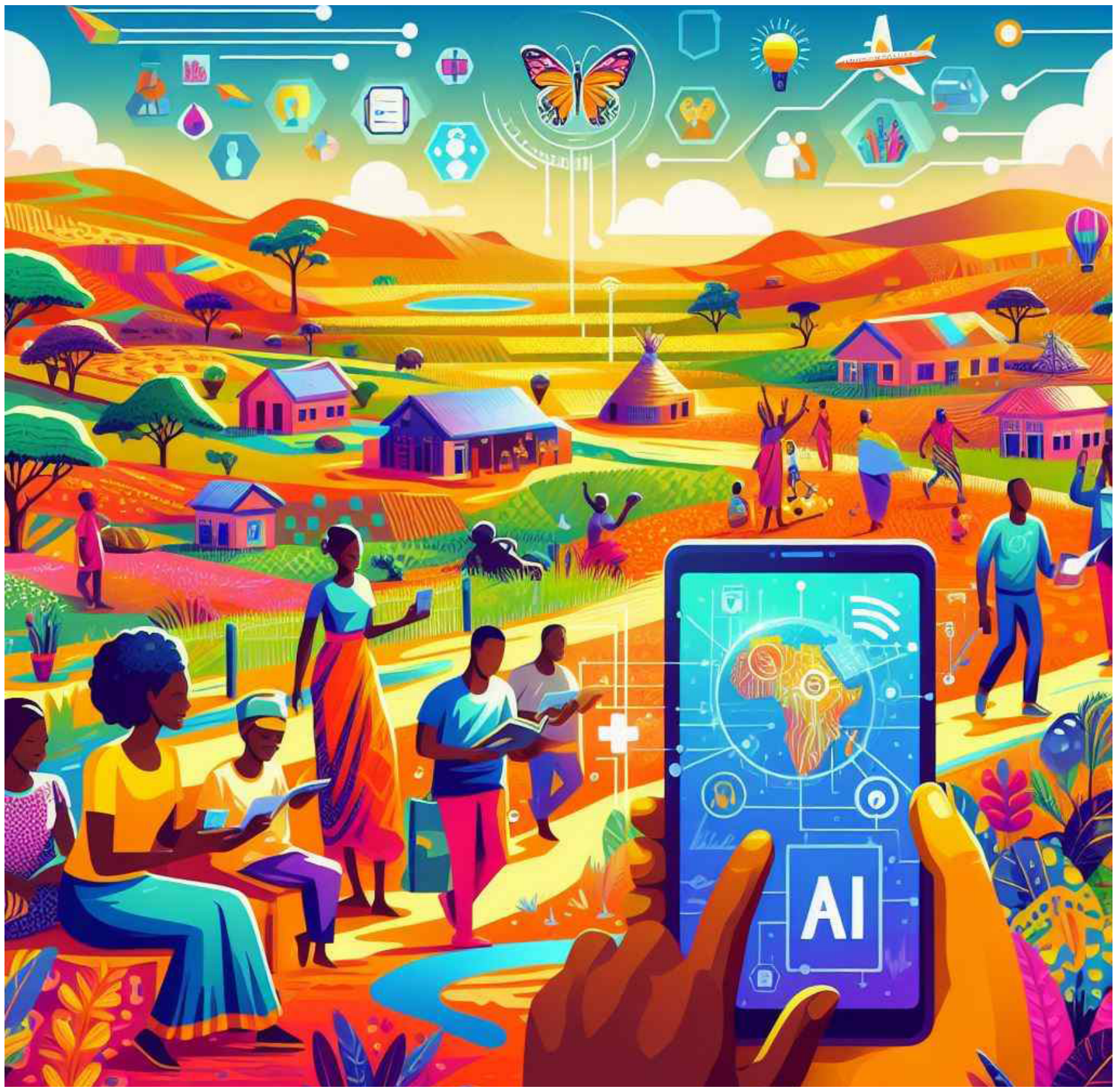

The impact of AI on the future that emerges will be a consequence of many things, including technological and policy decisions made today. Getting to a better future will require carefully designed policies and regulations that foster the development of AI while keeping the negative effects in check. To maximize the benefits and mitigate the macroeconomic risks related to Generative AI, leaders must also invest in digital infrastructure and human capital - including education initiatives - whilst ensuring that AI development is inclusive and tailored to the continent's unique needs and challenges. Addressing these issues is essential to ensure that AI acts as a catalyst for equitable and sustainable growth in Africa.

# Jobs, Skills and Labour Markets

## Background

Africa's rapidly growing population and vibrant tech ecosystem provide significant opportunities, particularly through the emergence of startups, tech hubs, and collaborative projects that drive digital transformation. The youthful energy and widespread availability of digital tools in sectors from agriculture to finance are catalyzing the continent's dynamism and can position Africa as a leader in technological innovation and sustainable development.

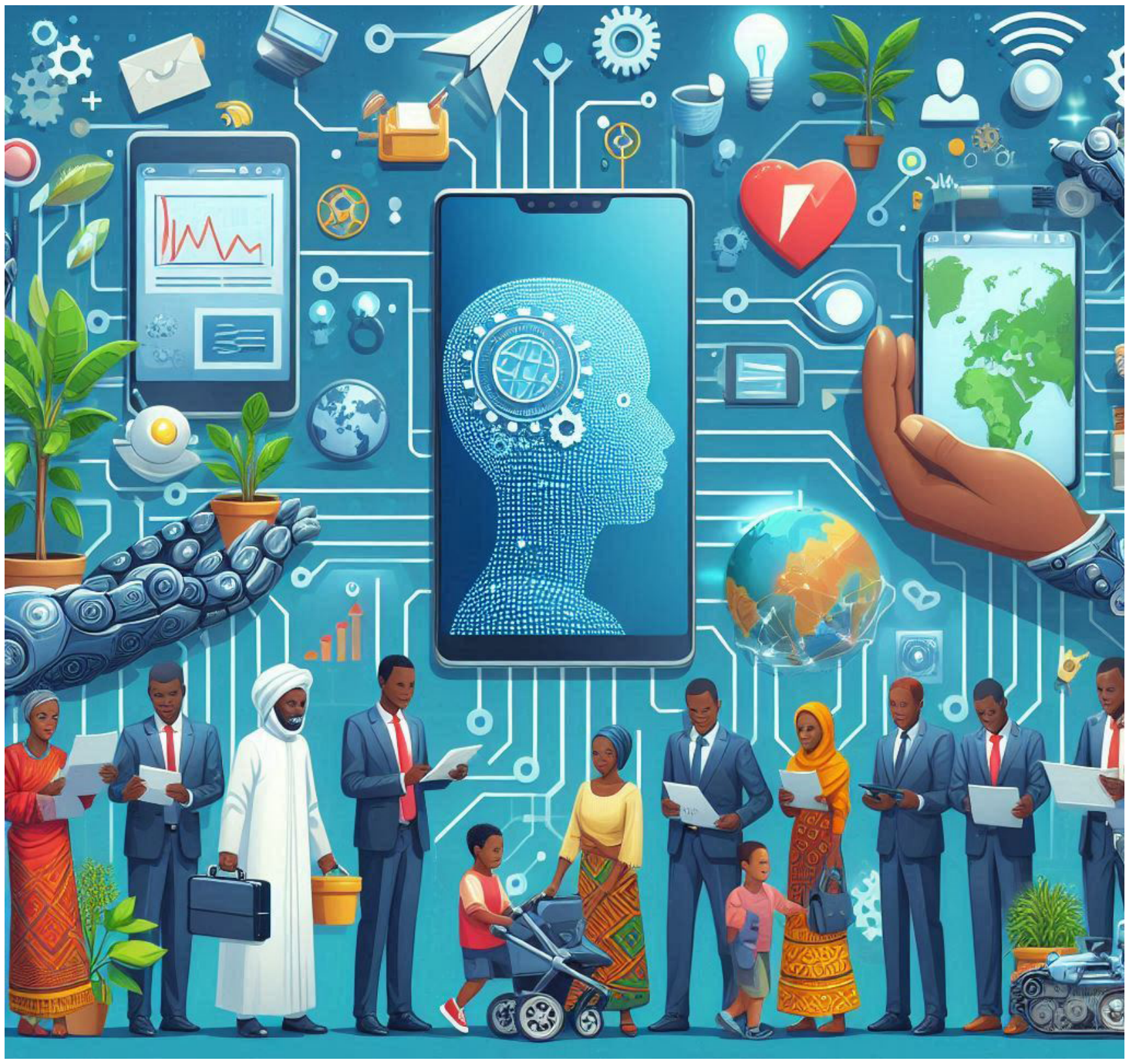

**The young demographic could provide a powerful driver for employment, development and economic growth. However, a recent report highlighted the challenge for African countries in creating jobs and equipping young people with the necessary skills for economic advancement[1]. Against this backdrop, advances in Generative AI are expected to be a double-edged sword. On one hand, there is potential to employ relevant technologies to markedly increase access to education, training and skilling and ensure African youth can forge ahead. On the other hand, current and expected disruptions in the labour markets could deepen the unemployment rates across the continent and entrench the already unsustainable social and economic inequalities. These concerns are grounded in past experiences of historic exclusions[2], which if not addressed are likely to continue to play out as Generative AI unfolds.**

**For example, out of the 29 countries in the Global Partnership on Artificial Intelligence, designed to: "guide the responsible development and use of artificial intelligence, grounded in human rights, inclusion, diversity, innovation and economic growth" only one African country – Senegal – is represented.**

**This section explores the unexpected consequences of labour market disruptions in Africa and dialogues around Generative AI and skilling. It ends with some recommendations for building a positive future.**

## Unexpected Consequences of Labour Market Disruptions in Africa

The techno-optimist approach views labour disruption as necessary for the advancement of economies and the creation of opportunities for new, previously inconceivable career pathways. The promise of Generative AI is to accelerate the speed and magnitude of labour transitions, with likely disruptions to many industries. Several reports have provided insights on jobs most likely to be disrupted by AI. These have largely focused on developed markets, and very select countries in Africa (mostly South Africa)[3 4 5 6].

The IMF working paper on AI Occupational Exposure[7] generally indicates lower exposure in emerging markets (Brazil and South Africa), with high skilled workers with 'higher cognitive-based tasks' being particularly at risk. Emerging markets with a larger share of workers in agriculture and the informal sector tend to have lower baseline exposure to AI. Thus, Generative AI is predicted to have a delayed impact on African economies (IMF report, 2024).

This generalized data masks nuances and disparities including gender, education, age and location. These nuances play a critical role in educational attainment, labor force participation and income distribution. Additionally, such reports fail to account for the aspirational nature of professional 'higher cognitive' jobs in emerging markets. There is a clear mismatch in the aspirations of Africa's young people and the predicted labour markets. According to a study of young people across 10 African countries, jobs that provided security e.g. in the public sector were considered most attractive while medium-skilled jobs in agriculture and manufacturing were the least attractive[8].

1 Rocca, C. and Schultes, I., 2020. Africa's youth: Action needed now to support the continent's greatest asset. Mo Ibrahim Foundation, pp.2020-08. https://mo.ibrahim.foundation/sites/default/files/2020-08/international-youth-day-research-brief.pdf
2 Fuchs, C. and Horak, E., 2008. Africa and the digital divide. Telematics and informatics, 25(2), pp.99-116.
3 World Economic Forum (2023). Jobs of Tomorrow: Large Language Models and Jobs. September. Available at: [https://www3.weforum.org/docs/WEF_Jobs_of_Tomorrow_Generative_AI_2023.pdf].
4 McKinsey Global Institute, 2018. Notes from the AI frontier: Modeling the impact of AI on the world economy.
5 Pizzinelli, C., Panton, A.J., Tavares, M.M.M., Cazzaniga, M. and Li, L., 2023. Labor market exposure to AI: Cross-country differences and distributional implications. International Monetary Fund.
6 Bonnet, F., Vanek, J. and Chen, M., 2019. Women and men in the informal economy: A statistical brief. International Labour Office, Geneva, 20.
7 Melina, G., Panton, A.J., Pizzinelli, C., Rockall, E. and Tavares, M.M., 2024. Gen-AI: Artificial Intelligence and the Future of Work. [https://www.developmentaid.org/api/frontend/cms/file/2024/01/SDNEA2024001-1.pdf]
8 Lorenceau, A., J. Rim and T. Savitki (2021), "Youth aspirations and the reality of jobs in Africa", OECD Development Policy Papers, No. 38, OECD Publishing, Paris, https://doi.org/10.1787/2d089001-en.

In addition, the ready availability of off-the-shelf generative AI tools for anyone to use anywhere, as long as they have an internet-enabled device and data, could enable a greater disruption of African markets than predicted. Although device and internet availability and affordability are still a challenge especially for rural populations. It is likely that disruption will therefore be skewed – potentially with big city based small businesses willing and able to adopt Generative AI tools, whilst businesses elsewhere are more likely to be excluded.

Nonetheless, barriers to access have been reduced compared to previous generations of AI technologies which typically required machine learning skills for implementation in workplaces. By comparison, even small and micro businesses can, and are, already deploying some generative AI tools and services such as ChatGPT, AI-enhanced search and image generation in the workplace.

While AI could herald new career opportunities in the future, the gaps during the transitions could prove difficult to navigate. The most critical priority presently is to address the increasing unemployment burden, and create employment opportunities at scale, to meet current and growing demand. In addition to creating new opportunities in the formal sector, are there ways in which generative AI could be used to boost the informal sector?

## Skilling, Upskilling, Reskilling

As demonstrated in the next section on Workers Perspectives, the biggest impact from Generative AI is likely to be found in information work which already has some level of digitization. This includes work in both the knowledge and frontline work sectors, and includes supporting communications, such as emails and presentations; creating sales and marketing materials, from posters to digital flyers; ideation, planning and so on. It is most likely to be used, at first, in the formal and semi-formal sectors. Over time, Generative AI may well become embedded in more diverse sectors and systems – such as agriculture, healthcare, financial inclusion – as more sector specific applications are created.

To be able to take advantage of even the off-the-shelf systems requires access to devices and data[9], knowledge of their existence and the skills to make use of them[10], as well as the ability to translate use into desired outcomes[11]. That is, all the factors in the traditional digital divide need to be bridged. As such, the existence and potential uses of such systems would need to be publicized, beyond the more connected city dwellers. In addition, the working population will need to understand how to effectively interact with Generative AI systems. Given the novelty and fast-changing nature of Generative AI technologies, this is currently a challenge for almost all populations globally, including office-based workers in large enterprises.

This all begs a set of questions: Who needs what skills to effectively harness the full power of Generative AI for Africa's development? How do we ensure that Generative AI works effectively for African contexts[12] so as not to compound existing systemic and colonial inequalities? And relatedly, who should be developing generative AI solutions for African contexts (see Section on Afro-Centric AI Platforms)? What institutions are required to provide these skills at-scale across the continent? What are the most effective training tools to achieve this?

Even as the development and deployment of Generative AI systems progress, an emergent value chain model can be used to explore the skills gaps in African markets. The value chain model adapted from a McKinsey report suggests that Africa's role might be relegated to the last mile of the value chain. Whilst it is not necessarily the case that Africa's role will be relegated to the last mile in the value chain, significant obstacles such as: cost, computing capacity, reliable power and a lack of investor confidence will need to be overcome for African countries to reap the full benefits of Generative AI. There are questions then to be asked about where the best play for African

9 Fong, E., Wellman, B., Kew, M. & Wilkes, R. Correlates of the digital divide: Individual, household and spatial variation. In Office of Learning Technologies, Human Resources Development (Citeseer, 2001).
10 Hargittai, E. Second-level digital divide: mapping differences in people's online skills. arXiv https://arxiv.org/ftp/cs/papers/0109/0109068.pdf (2001).
11 Van Deursen, A. J. & Van Dijk, J. A. The first-level digital divide shifts from inequalities in physical access to inequalities in material access. New Media Soc. 21, 354–375 (2019).
12 And who decides what "effectively" is industry, academia and governments are in these fast-moving markets[13]. Whatever the answer(s), it is as important to build skills across the spectrum, from how to deploy and use generative AI tools effectively at work, to how to build appropriate and innovative applications and technologies on top of these models, to the post-graduate skills of research and innovation in machine learning, natural language processing, human computer interaction, cybersecurity and systems to name but a few. Investing in this range of skills gives Africans the best opportunity to create dignified, appropriate jobs, to adapt AI sensitively to indigenous knowledge, to create new value chains and better AI systems which might reflect for example human-centred and community values. Such systems would add value globally and could counter typical tech-centric models of automation and deskilling.

Relatedly, we cannot ignore the vast spectrum of data work that is necessary to make these systems work. Currently, this is often poorly paid and under-valued[14]. An important part of creating dignified work globally would require reframing and rethinking this work as essential, skilled and valued labour.

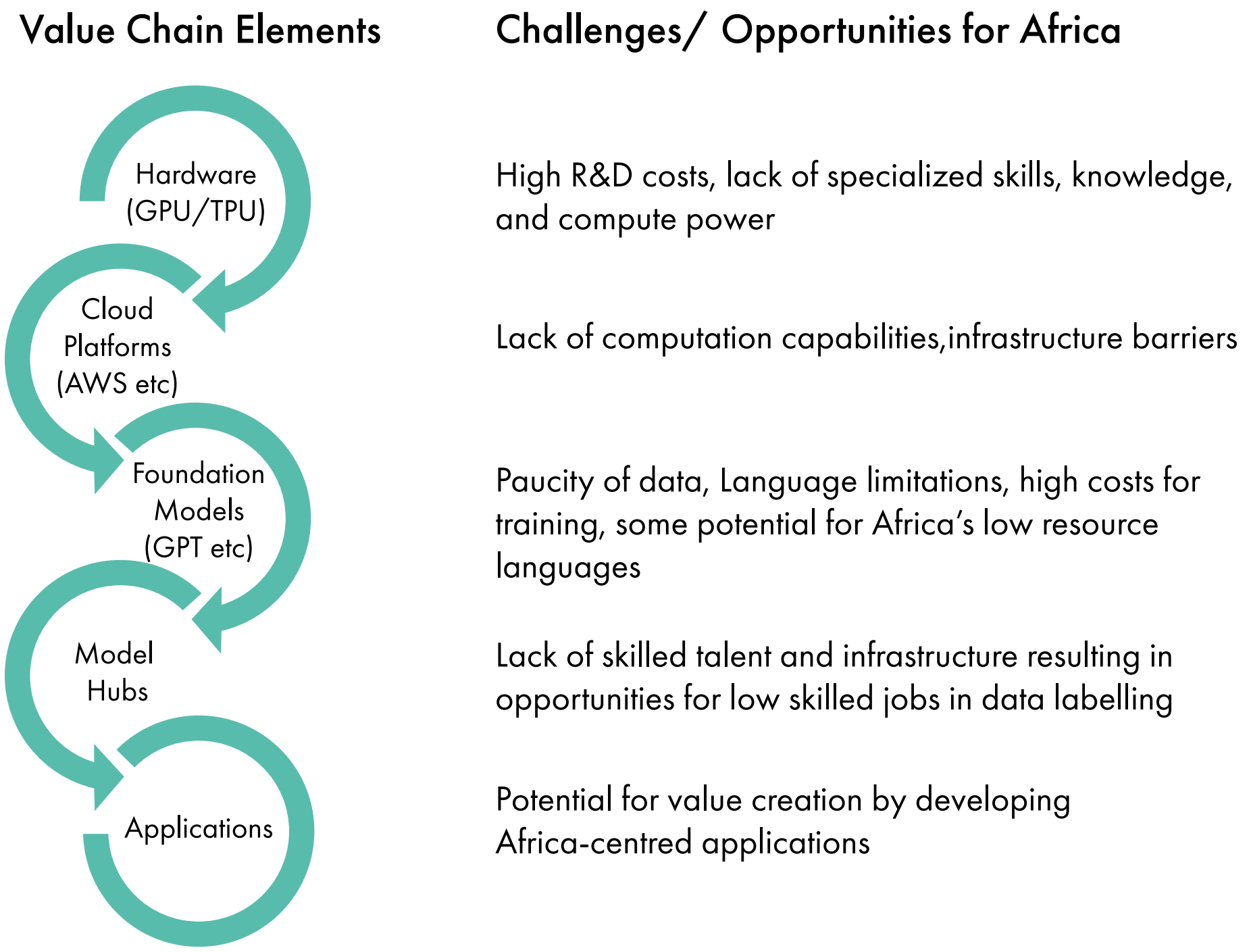

Globally, in the short term, various workforces are likely to require cycles of skilling, reskilling and upskilling to match the fast pace of Generative AI evolution and the rapid discovery of new use cases. This requires a rethinking of how we do 'on the job' learning, what career paths will look like and how any worker and workforce can keep pace during this rapid period of change. However, it is important to be careful about how we define and understand skilling[15]. As the ways in which Generative AI changes work and workplaces becomes clearer, the sets of new skills necessary for different jobs will also become clearer. Of course, the best technology fits into people's natural work practices, offering new opportunities and possibilities and transforms work more organically[16].

13 Questions which only they can answer

14 TIME. (2023). Exclusive: OpenAI Used Kenyan Workers on Less Than $2 Per Hour to Make ChatGPT Less Toxic.Available at: https://time.com/6247678/openai-chatgpt-kenya-workers/

15 Mutandiro, K. and Adeleke, D.I. (2024) 'African universities are failing to prepare tech graduates for jobs in AI,' Rest of World, 3 May. https://restofworld.org/2024/ai-skills-training-africa/.

16 The technical apparatus is, then, being made at home with the rest of our world. And that's a thing that's routinely being done, and it's the source of the failure of technocratic dreams that if only we introduced some fantastic new communication machine, the world will be transformed. Where what happens is that the object is made at home in the world that has whatever organisation it already has. -- Harvey Sacks (Lectures on Conversation Vol. 2., 548-9)

Nonetheless, the 'how to learn' skills are likely to become even more important than the 'what to learn'. Dialogic learning goals will be required for learners to conceptualise knowledge by asking meaningful questions while building collaborative and humanism skills[17].

Along the spectrum of digital literacy, a majority of the population will be expected to have enough technical literacy to interact with Generative AI systems, a significant proportion of the workforce will be required to have proficiency skills to meaningfully engage and shape the systems (e.g. business leaders and policy makers) and a proportion of experts needed to drive the development of Africa-centric Generative AI systems. To effectively harness and interact with Generative AI systems, non-technical and transferable skills including problem solving, creativity, adaptability, and critical thinking are essential. This necessitates a different approach to digital skills development.

Currently, only half of African countries have computer skills in their school curriculum, compared to a global average of 85%[18]. To put a global perspective on this, African countries currently score between 1.8 and 5 on the Digital Skills Gap Index, lower than the global average of 6[19]. National programs such as Rwanda's smart classroom initiatives provide models for scalable digital literacy programs[20].

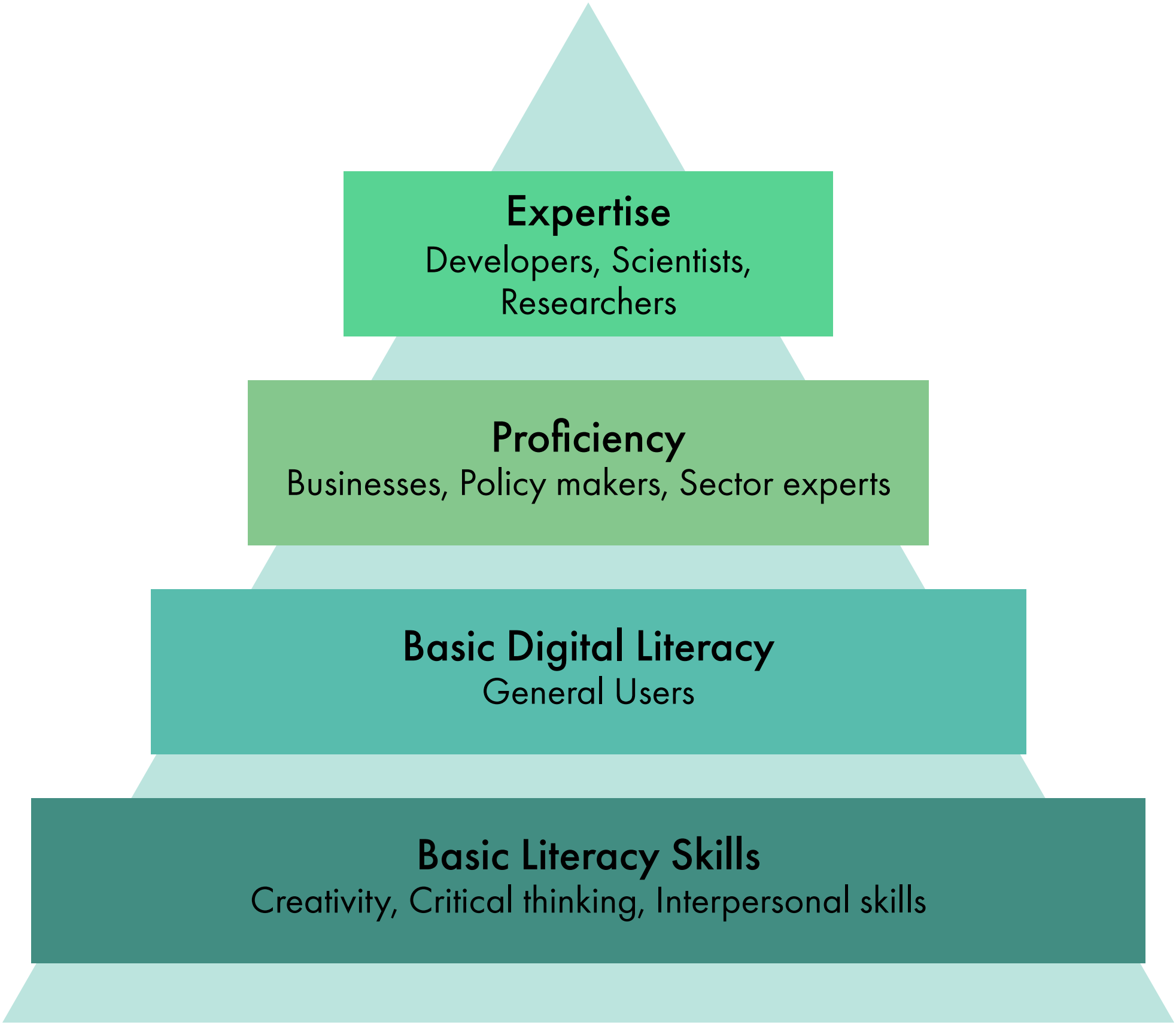

17 Anapey, G. M. (2023). Exploring Ai and dialogic education outcomes from learning sciences perspective. In L. G. Amoah, Examining the rapid advance of digital technology in Africa (pp. 147-162). DOI: 10.4018/978-1-6684-9962-7: IGI Global.

18 Kandri, S.-E. (2024) 'Africa's future is bright—and digital,' World Bank Blogs, 16 March. https://blogs.worldbank.org/en/digital-development/africas-future-bright-and-digital.

19 Dupoux, P. et al. (2022) Africa's opportunity in digital skills and climate analytics. https://www.bcg.com/publications/2022/africas-opportunity-in-digital-skills

20 Welcome | Smart class. https://www.smartclass.rw/

# Takeaways and Recommendations

Recommendations drawn from the deliberations can been categorised into four parts.

## 1. Africa-led and Africa-owned Research agenda

In-depth research on the scale of Generative AI disruption of work across various sectors of African economies is critical. Beyond the usual generalizations, a critical lens needs to be applied to understand the nuances of the repercussions of Generative AI to Africa's unique social and economic contexts.
Fully understanding and mapping the existing expertise and skills of African youth at sub-national, national and regional level can help identify whether and where there are skills gaps[21].

## 2. Skilling for the entire sector: from building AI to working with it

**Building and deploying AI.** There are a variety of factors which could limit the participation of African countries, academia and industry in developing new large language and multi-modal models[22]. This lack of representation, compounded by the data divide (see 'Workers Perspectives' section) already results in instances of biased and exclusive AI outputs from AI systems when faced with African contexts and languages. It is therefore vital that all organisations which develop and apply Generative AI models within African countries, do so with a sensitivity to the varied contexts that exist. Fostering research (including humanities and social sciences), technology, human-centred design and engineering skills across the entire AI value chain in Africa would enable the creation of more inclusive and representative AI technologies. It is important to stress that social science, humanities, and human- and community-centred design are as important in building AI in and for Africa as anywhere else. Without including these disciplines in a central role for AI design the majority of challenges around inclusiveness and representativeness of AI are unlikely to be addressed.

**Working with AI.** In addition, skills for working with AI and AI-assisted jobs are needed, alongside the people-centric, transferrable skills that have always been essential for success in the workplace. Not specific to African workers, but required globally, these include creativity, agile thinking, communication, integrity and judgement. These skills are likely to remain relevant for jobs involving AI. While these universal skills can be taught experientially, they require alignment with community values and culture.

**Education.** Globally, adoption of enquiry-based pedagogical approaches in education institutions and integration of AI across all subjects and learning outcomes will be critical in creating an adaptable workforce. Nurturing higher education and research will be a central part of ensuring a highly skilled workforce ready to participate in all parts of the AI value chain. Partnerships with global research and education institutions is one avenue for specialized training for Africa's AI experts, but it is equally, if not more, important to nurture the growing AI research centres in academic institutions across the continent. Specific interventions and incentives also need to be put in place to attract and retain Africa's tech talent to serve Africa's interests.

21 Adams, R., Alayande, A., Brey, Z. et al. A new research agenda for African generative AI. Nat Hum Behav 7, 1839–1841 (2023). https://doi.org/10.1038/s41562-023-01735-1
22 This is not unique to Africa, and equally impacts Europe, Global S

## 3. Preparation and adaptation of labour markets

Given the widespread reports of the deplorable working conditions of low-wage data annotators across the continent, it is evident that the AI sector is likely to replicate existing extractive and often unethical labour practices in Africa[23]. It is important that Africa is not relegated to being the hidden army of low skilled AI data workers[24] preparing and annotating training datasets. Even with exceptional training and capacity development, the unpredictability and uncertainty of AI disruption renders millions of currently stable employees at risk. This race to the bottom can only harm African workers and needs to be remedied through global voluntary and statutory policy action[25]. Because of the global labour market[26] that all digital work finds itself in, any attempt to improve standards in one country can simply cause jobs to flow to another country with weaker protections. As such, there is a need for global standards that firms in production networks adhere to. African markets must also coordinate their protection mechanisms and policies to avoid a race to the bottom. The ILO report on decent work in the platform economy can provide some guidance in this respect[27]. Voluntary Mechanisms such as the Fairwork AI Principles, modelled off similar and successful approaches in the garment industry, are another mechanism which can be used. Lead firms in the Global North can embed the principles into their supplier agreements: thus, signaling to all suppliers that they must abide by the same set of minimum standards. This approach has been shown to push up standards in companies which commit to it[28] and is a promising complementary approach to legislation.

**The involvement of youth, community leaders, academics, and business leaders are critical in developing inclusive and relevant AI policies for Africa**

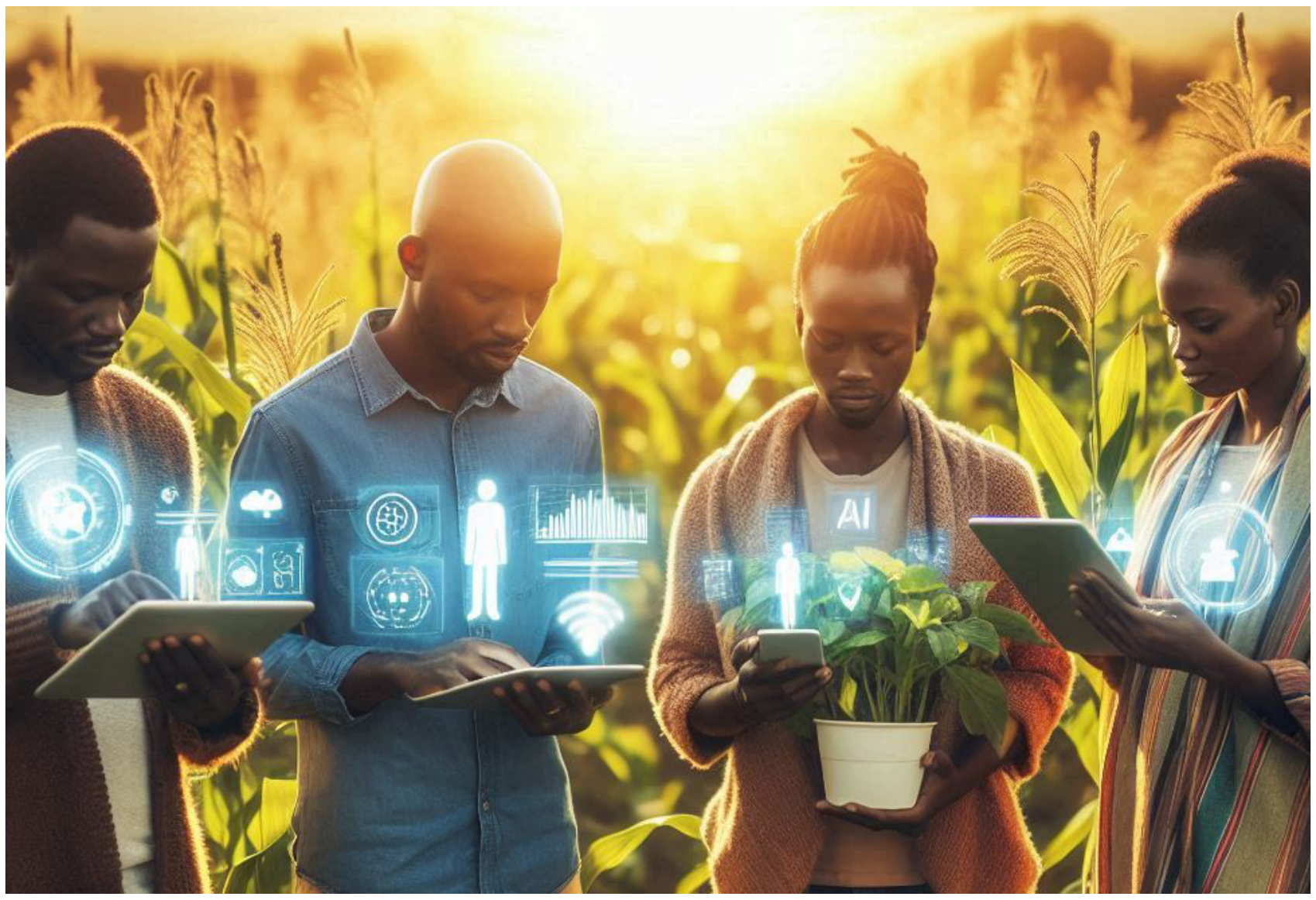

23 Muldoon, J., Cant, C., Graham, M. and Ustek Spilda, F., 2023. The poverty of ethical AI: impact sourcing and AI supply chains. AI & SOCIETY, pp.1-15.https://link.springer.com/article/10.1007/s00146-023-01824-9

24 Anwar, M.A. and Graham, M., 2020. Digital labour at economic margins: African workers and the global information economy. Review of African Political Economy, 47(163), pp.95-105.https://www.tandfonline.com/doi/abs/10.1080/03056244.2020.1728243?journalCode=crea20

25 Anwar, M.A. and Graham, M., 2022. The Digital Continent: placing Africa in planetary networks of work (p. 288). Oxford University Press. https://global.oup.com/academic/product/the-digital-continent-9780198840800?cc=gb&lang=en

26 Graham, Mark, and Mohammad Amir Anwar. "The global gig economy: Toward a planetary labor market." In The digital transformation of labor, pp. 213-234. Routledge, 2019.

27 International Labour Conference (2025) Realizing decent work in the platform economy, International Labour Conference. report. International Labour Office, p. 3. https://www.ilo.org/wcmsp5/groups/public/---ed_norm/---relconf/documents/meetingdocument/wcms_909906.pdf.

28 Muldoon, J., Cant, C., Graham, M. et al. The poverty of ethical AI: impact sourcing and AI supply chains. AI & Soc (2023). https://doi.org/10.1007/s00146-023-01824-9

### 4. Evolving and inclusive AI policy frameworks

African Union digital strategies for education, agriculture, and health provide a foundation on which to reframe and build Africa's AI strategy and policy frameworks. While the African Union strategy for AI is underway, national and regional frameworks need to be developed and implemented to provide relevant skills and protect workers from exploitation. In the absence of these policies, competitive market dynamics of the global AI supply chain are likely to result in price wars, in particular around labour, making workers susceptible to exploitation and abuse.

Existing inequalities in education outcomes need to be redressed to create inclusive AI literacy skills development. Specific focus on gender and disability inclusion is paramount in creating AI systems that are representative of the general population.

As evidenced in the workshop, the involvement of youth, community leaders, academics, and business leaders are critical in developing inclusive and relevant AI policies for Africa. This requires a more agile consultative policy formulation process with sufficient scope for improvement as the Generative AI space evolves.

# African Workers' Perspectives on Generative AI

## Introduction

In this section we examine workers' perspectives around the use of Generative AI, with a focus on the intersection of Generative AI with language, context, culture, and data. African workers encompass a wide variety of people in a range of jobs from frontline work to information work. In Africa, 85.8 percent of employment is informal[1], and the International Labor Organization (ILO) estimates that 95% of African youth ages 15-24 work in an informal setting[2]. Street vendors, taxi drivers, hairdressers, metal workers and repair shops are some examples of the urban informal sector[3]. Entrepreneurialism is common and many workers have multiple jobs – or side hustles – including professionals in the formal sector. Within the start-up ecosystem and the formal sector of many African countries, there is a well-developed knowledge economy with strong linkages to global capital. Governments across Africa's 54 African Union recognized countries are a significant employer with armies of workers of various capabilities spanning many disciplines. Finally, a large segment of workers who exist outside or marginally within the market economy exist in academia, civil society and across homes all across the continent.

To use Generative AI, workers must have access to devices – such as smartphones or computers - and internet/data; they must know about the tools and have the skills to use them in their work. One way of achieving connectivity is though digital hubs to which workers can be communally linked. For example, in Kenya, these range from simple hotspots in Nairobi, to village polytechnics in Kenya's 1440 wards[4]. They can also be found in cyber cafes that dot rural and urban centers across the continent.

Whilst the barrier to access for general purpose Generative AI tools is relatively low, it still excludes

1 More than 60 per cent of the world's employed population are in the informal economy (2024). https://www.ilo.org/resource/news/more-60-cent-worlds-employed-population-are-informal-economy

2 Bonnet, F., Vanek, J. and Chen, M., 2019. Women and men in the informal economy: A statistical brief. International Labour Office, Geneva, 20.

3 Karlen, R., Rougeaux, S. and De Silva, S.J. (2024) 'From Market Stalls to Mechanic Shops: Better Jobs for Côte d'Ivoire's Urban Youth,' World Bank Blogs, 16 March. https://blogs.worldbank.org/en/africacan/market-stalls-mechanic-shops-better-jobs-cote-divoires-urban-youth.

4 Nyamori, M. (2023) 'State to establish 1,450 village digital hubs, 25,000 WiFi hotspots nationwide,' Nation, 11 February. https://nation.africa/kenya/news/state-to-establish-1-450-village-digital-hubs-25-000-wifi-hotspots-nationwide-4120356.

the many workers who do not have ready access to data[5]. Outside of North Africa, African consumers pay disproportionately higher prices for mobile phone data, and five out of the top ten most expensive countries are African[6].

Indeed, the high price of data limits mobile internet use even for those with an internet enabled smartphone[7]. Even in countries where mobile data is becoming more affordable, broadband connectivity remains pricy[8]. Many of general-purpose Generative AI tools are free to use to some extent – but more extensive business use or tailored tools will often require paid subscriptions, which are likely to be out of reach for many workers. Whilst, we have seen adoption of generative AI tools in Small and Medium Businesses in Kenya and Nigeria, the extent of penetration of these generative AI tools across Africa is not yet clear.

**African workers are highly diverse, from urban to rural, frontline to information workers, start-ups to enterprises, and with up to 85% working in the informal sector**

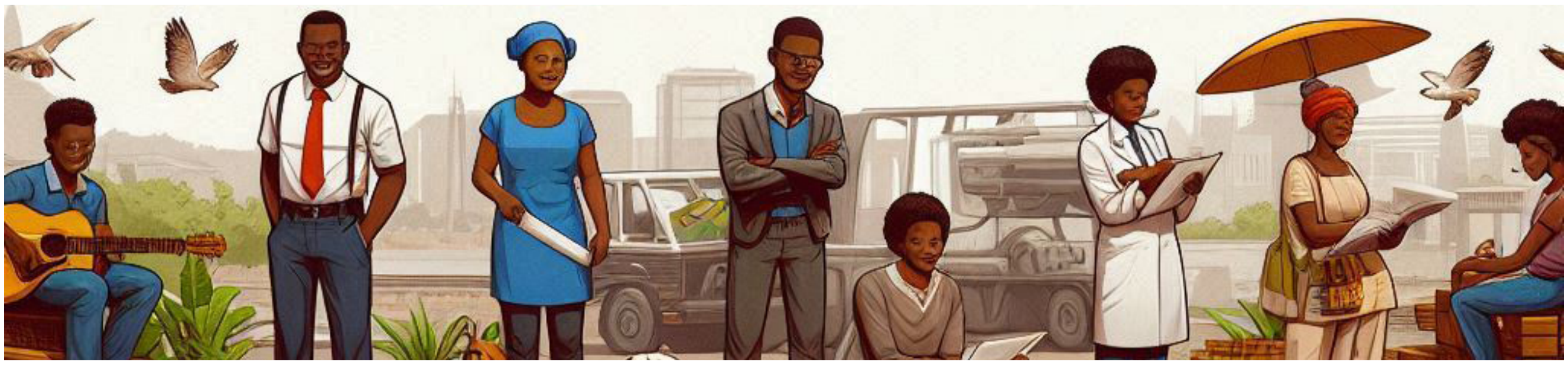

Given the above, we can loosely classify the accessibility of Generative AI tools to workers:

- Generative AI is largely not accessible to workers who do not know about it or how to access it; workers who do not have the confidence to try out new technologies – or are skeptical about their benefits over their potential risks; who do not have smartphones or data; who have limited literacy; who are working for companies with limited funds for technology adoption or are working in areas with less developed technological infrastructure (i.e. rural areas).
- Generative AI may currently have limited use for workers who are running frontline businesses or doing frontline work where the vast majority of their work is physical and does not include digital work; who are doing well defined work, such as teachers in many schools, where there is little

5 GSMA notes that "While 94% of the global population are now covered by mobile broadband, 450 million people still remain uncovered, with the vast majority (47%) in Sub-Saharan Africa." https://www.gsma.com/r/wp-content/uploads/2021/09/The-State-of-Mobile-Internet-Connectivity-Report-2021.pdf; GSMA also estimates that 348M people in Africa don't have access to an internet-enabled phone (page 15) https://www.gsma.com/solutions-and-impact/connectivity-for-good/mobile-for-development/wp-content/uploads/2022/04/Making-internet-enabled-phones-more-affordable-in-low-and-middle-income-countries.pdf
6 As young Africans push to be online, data cost stands in the way (2022). https://www.weforum.org/agenda/2022/06/as-young-africans-push-to-be-online-data-cost-stands-in-the-way/.
7 Delaporte, A. et al. (2022) Making internet-enabled phones more affordable in low- and middle-income countries, GSMA. https://www.gsma.com/solutions-and-impact/connectivity-for-good/mobile-for-development/wp-content/uploads/2022/04/Making-internet-enabled-phones-more-affordable-in-low-and-middle-income-countries.pdf. Pg36.
8 Kariuki, N. (2023) Breaking down Kenya's broadband costs: A comparative analysis. https://www.linkedin.com/pulse/breaking-down-kenyas-broadband-costs-comparative-analysis-kariuki/.

opportunity for self-expression and deviation from the curriculum; who work in sectors that are low-tech such as construction, agriculture, and small-scale manufacturing. Although as new tools and interaction mechanisms develop, one could imagine greater adoption in the future, as customized tools are built for different sectors.

- Generative AI is likely to have most impact for workers who are already using technology in their job (computer, tablet or phone); who are doing a lot of clerical work; in the more desk-based professions – or where industry sectors develop customised tools e.g. for healthcare professionals; for workers doing design work, graphic work, sales and marketing materials. These workers can already benefit from existing tools, although there is much room for improvement in tools which can generate production quality or customer-ready content.

At least initially, the biggest impact of Generative AI is likely to be on information work – especially where workers are working on digital tools at a desk. Types of work include content creation and management; research and data analysis; library and archival services e.g. digitisation of collections to make them more accessible and to preserve them; educational resources, especially in higher education; information dissemination e.g. to help draft public announcements and information; and legal and statutory requirements e.g. document review. In several sectors, Generative AI can help with marketing and customer engagement using targeted campaigns, operational efficiency through automation of routine tasks, and customer support services that are available 24/7.

In the rest of this section, we discuss adoption and skills; what an ideal future might be, cultural alignment and clashes and end with some key takeaways.

## Adoption and Skills

Generative AI's adoption in Africa reveals the coexistence of two contrasting perspectives. On the one hand, its rapid uptake and user-friendliness are evident. Many African businesses and individuals are quickly embracing Generative AI, drawn by its clear benefits, access to free and low-cost versions, and the simple requirements of a device and internet access. This accessibility has spurred widespread experimentation and application in work settings.

However, alongside this enthusiasm, there's apprehension, including fears of job losses and the uncertainty of change[1] [2]. Such worries are common globally and are fueled by debates about General AI and hypothetical AI sentience. When combined with locally experienced negative impacts such as the impact on creators of Generative AI in advertising[3] this can lead to pessimism. The historical contexts of Western relations and the current limitations of Generative AI in African contexts are likely to further compound this. These dynamics raise critical questions about the need for Africa-Centric AI, its potential form, and its relationship with Western AI models—a topic we revisit.

1 O'Neill, J. (2024) 'How AI can revolutionise Africa's labour landscape.,' Business Daily, 16 April. https://www.businessdailyafrica.com/bd/lifestyle/workplace/how-ai-can-revolutionise-africa-s-labour-landscape--4592556.
2 Owino, V. (2023) 'Jobs that will survive AI, and those that won't,' The East African, 22 September. https://www.theeastafrican.co.ke/tea/science-health/jobs-that-will-survive-ai-and-those-that-won-t-4377378.
3 Agencies (2024) Kenyan marketers embrace generative AI as its use spreads globally. https://www.capitalfm.co.ke/business/2024/01/kenyan-marketers-embrace-generative-ai-as-its-use-spreads-globally/.

In terms of skills and the African worker, four key elements emerge:

1. **Preparedness**: Governments, educational bodies, and employers must be agile in reskilling workers affected by technological shifts. Predicting the exact impact of Generative AI on jobs is challenging, however, it is likely to transform many information work jobs, from journalists to legal clerks. Exactly how this will play out remains to be seen, but it's vital to stay responsive to these changes. Further, it is vitally important for both the quality and well-being of society, as well as the individual workers, that these transformations improve the quality of the work produced and support and enhance the creativity and value of workers, rather than engaging in a race to the bottom.

2. **Skill Development**: People need the skills, knowledge, and access to leverage Generative AI in their work and careers. This includes understanding what tools exist, how to apply them effectively, and critically evaluating their output. We are already seeing evidence that some Small and Medium Businesses feel compelled to use these tools so that others who do use them do not get an advantage over them. Given the tools' propensity for fabrication, knowing how to evaluate and appropriately deploy their output will become an important new business skill. Overseeing machine output is challenging, and strategies to recognize risky AI outputs are crucial[4]. Providing workers with the means to identify potentially risky model output is likely to be an important part of the design and deployment of Generative AI tools.

3. **Local AI Leadership**: For Africa to significantly contribute to the AI economy, it's essential to cultivate African talent in AI research, innovation, and design, as well as policy and governance. This requires top-notch education and research opportunities within Africa as well as access to global opportunities. A strong development of multi-disciplinary programs will become increasingly important, covering machine learning, human-computer interaction and AI ethics and including both computer and social scientists. Currently, underfunded AI research and a lack of postgraduate opportunities in many African countries hinder this development[5]. A shift towards equitable AI development demands more public, as well as private

## In an ideal future, generative AI might provide the following benefits

In the best future, Generative AI could offer a transformative path for African workers, enhancing their capabilities and fostering a brighter future:

**Enhanced Summarization and Synthesis**: Generative AI could streamline the rewriting and repurposing of material, enabling workers to focus on creative aspects while reducing monotonous tasks.

**Revolutionizing Reporting**: By automating tedious aspects, workers could be empowered to dedicate more time to imaginative and innovative elements of reporting.

**Interactive and Comprehensive**: By interacting across various information sources, the user experience could be smoother.

**Personalized and Adaptable**: If tailored to individual workers, learning from interactions and becoming personalized, whilst respecting privacy, Generative AI might enhance each worker's unique skills.

**Community-Centric Assistant**: Reimagining Generative AI as a community-focused tool, it could support collaborative work and communal development.

**Personal Assistant for Skill Enhancement**: Generative AI could offer constructive feedback and advice to improve work quality and avoid pitfalls.

**Entrepreneurial Aid**: Generative AI could assist in risk assessment and data analysis, empowering entrepreneurs in their ventures.

**Empowering Informal Sector**: Tailored Generative AI tools might elevate the capabilities of informal entrepreneurs, providing customized assistance for their unique needs.

**Promoting Equal Opportunities**: Generative AI could be used to facilitate diverse perspectives, reduce biases and foster a more equitable working environment.

**Language and Cultural Support**: It could offer translation and cultural comprehension support, enabling work in multiple languages without sacrificing quality.

**Ideation and Creativity**: Generative AI could help people quickly come up with new ideas.

**Augmentation, Not Automation**: Workers should be augmented by AI's functionalities, enhancing their roles rather than replacing them.

**Focus on Human Skills**: Generative AI might free up time for workers to concentrate on relationships, social skills, creativity, and imagination, essential aspects of fulfilling work.

4 Sellen, Abigail, and Eric Horvitz. "The rise of the AI Co-Pilot: Lessons for design from aviation and beyond." arXiv preprint arXiv:2311.14713 (2023).

5 Although see Africa's postdoc workforce is on the rise — but at what cost? (nature.com) for a perspective on some of the challenges of building a highly skilled research workforce

4. **Generative AI as a Skill Enhancer**: Generative AI offers potential as a tool for skill enhancement, enabling people to perform more creative and effective work than possible with AI or human effort alone.

## African Workers: An Ideal Future

In envisioning an ideal future for African workers, several key elements stand out:

1. **Collaboration with AI**: The core principle should be a synergistic relationship between workers and AI, where Generative AI flexibly assists workers in tasks they choose.

2. **Centering African Perspectives**: Development and implementation of Generative AI must prioritize African workers' perspectives, alongside organizational needs. This calls for robust negotiation channels, such as unions and work councils for the formal sector, plus the facilitation of worker collectives for the informal sector. Flexibility in tool deployment and personalization can ensure both collective and individual worker preferences are respected. The goal is to tailor AI support to the varying, situated needs of workers. For example, whilst African workers want support in reducing the mundane parts of their job, it is not necessarily desirable to have to be creative or challenged all day every day. Similarly, relationships are typically prioritised in African business[6], finding the right ways to use Generative AI to support rather than erode interpersonal communication will be crucial. The importance of tool flexibility and workers situated control over deployment is paramount.

3. **Fulfilling Creative Potential**: Empowering African workers to reach their creative heights will benefit the global community. This includes achieving a healthy work-life balance, where productivity gains from Generative AI for example reduce work hours without affecting pay, allowing more time for community, family, and work that brings societal value.

4. **Valuing Social Contributions**: The market economy often overlooks or undervalues vital societal roles like caregiving, education, art and environmental work. A future where such contributions are recognized and valued, potentially through mechanisms like a basic income, will enable more fulfilling lives.

5. **Work and Wellbeing**: Ensuring the wellbeing of African workers is paramount, independent of work opportunities. This means creating environments where people contribute meaningfully to society while enjoying life.

6. **Bridging Global Skill Gaps**: African workers should be able to fill skill gaps globally, earning fair wages without leaving their communities. One mechanism for this could be to create a global skills market, similar to the global energy market. At village-level, infrastructure (transport, education, healthcare) needs to be enhanced to foster sustainable, flourishing communities, benefiting both rural and urban areas by reducing migration pressures.

In this envisioned future, African workers will leverage Generative AI to enhance their roles, contribute globally, and lead more balanced, meaningful lives.

**Generative AI must be a flexible assistant, respecting and prioritizing the perspectives and needs of African workers, nurturing relationships, and enhancing societal roles in care, education, art and conservation to support their wellbeing and work-life balance**

6 Awori, K., Allela, M.A., Nyairo, S., Maina, S.C. and O'Neill, J., 2022. " It's only when somebody says a tool worked for them that I believe it will work for me": Socio-tecture as a lens for Digital Transformation. Proceedings of the ACM on Human-Computer Interaction, 6(CSCW2), pp.1-24.

## African Perspectives on Generative AI: Cultural and Societal Alignment and Clashes

The performance of generative AI models depends on the amount and quality of training data. However, most of the training data for existing generative AI models is sourced from the predominantly English-speaking Global North[7] and is hardly representative of African social and cultural realities. This can impact the effectiveness and reliability of AI applications for African workers. To understand this better, let's examine the representation of 1) language, 2) context and culture, and 3) data in generative AI models.

### Language

Africa has a rich and diverse linguistic landscape, which is not necessarily well represented in the training data of Generative AI. For example, GPT4 was trained on languages including Swahili, Afrikaans, Kinyarwanda, and Igbo and its performance in such languages is improving all the time, however performance on other less well represented African languages is likely to be more limited. Studies using traditional NLP benchmarks have shown that fully supervised models outperform generative models on African languages, especially in tasks like named entity recognition[8] [9]. However, whilst traditional NLP benchmarking studies such as Ojo's are important for highlighting the genuine need to improve the African language performance of Generative AI, they do not tell the full story. Many Africans speak multiple languages, both local and colonial, and in doing so, codemixing and local terminology is common. This presents a major challenge for traditional Natural Language Processing (NLP). Generative AI, particularly GPT4, has advanced NLP capabilities, being much better able to handle codemixing and informal language. For example, GPT4 effectively processed a Swahili-English-Sheng dataset, outperforming other traditional multilingual models without needing fine-tuning[10]. However, its reliability varies, especially with less familiar languages like Sheng.

Whilst these advances are certainly promising for African workers, there's still much to be done to achieve parity with high-resource languages[11]. Moreover, voice interfaces, preferred in many African contexts due to oral culture and varying literacy levels, lag behind. Current speech-to-text models struggle with multilingualism and accents, creating barriers to broader Generative AI application in the African workplace.

**There are over 3,000 languages spoken across Africa and many Africans speak multiple languages, both local and colonial, and in doing so, codemixing and local terminology is common**

### Context and Culture

African culture and context are also notably underrepresented in Generative AI training data. This is compounded by the fact that many sources of African data are undigitized or not available online. Often, the data about Africa that is included reflects external viewpoints, stemming from development agencies or foreign universities, rather than authentic African perspectives. This can lead to a distorted representation of African culture and context in

7 https://www.dataprovenance.org/
8 Ojo, J., Ogueji, K., Stenetorp, P. and Adelani, D.I., 2023. How good are Large Language Models on African Languages?. arXiv preprint arXiv:2311.07978.
9 Ahuja, K., Diddee, H., Hada, R., Ochieng, M., Ramesh, K., Jain, P., Nambi, A., Ganu, T., Segal, S., Axmed, M. and Bali, K., 2023. Mega: Multilingual evaluation of generative ai. arXiv preprint arXiv:2303.12528.
10 Ochieng, M., Gumma, V., Sitaram, S., Wang, J., Ronen, K., Bali, K. and O'Neill, J., 2024. Beyond Metrics: Evaluating LLMs' Effectiveness in Culturally Nuanced, Low-Resource Real-World Scenarios. arXiv preprint arXiv:2406.00343.
11 Ahuja, S., Aggarwal, D., Gumma, V., Watts, I., Sathe, A., Ochieng, M., Hada, R., Jain, P., Axmed, M., Bali, K. and Sitaram, S., 2023. MEGAVERSE: benchmarking large language models across languages, modalities, models and tasks. arXiv preprint arXiv:2311.07463.

AI models, being about Africa rather than from Africa. Representative data sets created and curated by Africans to reflect current realities on the continent are needed if Generative AI is to work well for Africans.

However, this brings attention to important questions about data governance and policy. Current Generative AI models are trained on vast sets of publicly available data, without clear guidelines on attribution and recompense. Africa has long been party to extractive industries – typically extracting raw materials from the continent – new models of data use and governance are needed if Generative AI is not to be the next extractive industry. Equitable Africa-centric Generative AI needs to consider not just what data is collected and curated, by whom, but also how it is used and the knowledge within recognized.

The predominance of an oral tradition in many African communities further complicates questions of data inclusion. Indigenous knowledge and local data, integral to African heritage, often remain undocumented and inaccessible for Generative AI training and fine-tuning. This omission risks excluding a wealth of African knowledge from AI applications, impacting both model relevance and cultural knowledge preservation. There is a clear opportunity here for voice and audio data sets as training data for Africa-centric Generative AI – which in itself could create work for Africans in the Generative AI pipeline. Again, if indigenous and oral data is to be represented and included in Generative AI, questions about data ownership, governance, attribution and compensation become central. New data sovereignty models are already being created, such as TeHiku Media's model which aims to give indigenous communities control over their own data[12].

The limited representation of African data in current models has consequences for users and use cases. Early research indicates underperformance of Generative AI for Small and Medium Businesses, in Kenya and Nigeria. With failures spanning text, image and voice generation tasks as well as speech-to-text[13]. The lack of representation in the training data is likely to have far-reaching impacts for African workers and highlights a critical concern: current generative AI models are "really Americanized", adversely affecting their utility for African businesses. Addressing these representation gaps is essential for creating AI tools that are truly beneficial and relevant for African business, however, as discussed, this requires innovation and policy to ensure equitable data ecosystems.

## Data

The underperformance of Generative AI models in African contexts is largely due to the amount and type of African data in the training sets, and biases in the various data related processes such as labelling and reinforcement learning. This raises some important questions:

1. Should representative African data be integrated into existing models, which are built and trained in the Global North?

2. Or should Africa-centric models be built and trained in Africa?

The African Union Data Policy framework[14] recognizes data as a strategic asset essential for policymaking, innovation and creating entrepreneurial opportunities. The report centres data justice, emphasizing the need for equitable and just outcomes in data governance, including the need for equitable inclusion in the data economy, and the balancing of individual and broader social and economic rights. It advocates for positive regulation to create an environment for effective participation in the digital economy and the need to build institutional capacity within states to enable this.

Considering the African Union Data Policy framework and its centering data justice, it is important to speak to those tensions around what models should be built and by whom. For both approaches, consideration needs to

12 Hao, K., 2022. A new vision of artificial intelligence for the people. MIT Technology Review. https://pulitzercenter.org/stories/new-vision-artificial-intelligence-people
13 Keynote: Building Globally Equitable AI - Microsoft Research (2024). https://www.microsoft.com/en-us/research/quarterly-brief/jun-2024-brief/articles/keynote-building-globally-equitable-ai/.
14 African Union [AU] (2022) The AU Data Policy Framework, African Union. EX.CL/ Dec.1144(XL). https://au.int/sites/default/files/documents/42078-doc-AU-DATA-POLICY-FRAMEWORK-ENG1.pdf (Accessed: June 7, 2024).

be paid as to

- Who will generate high-quality African datasets?
- How can we ensure data work is dignified work?
- What might equitable data ecosystems consist of?

The current Generative AI models rely heavily on unattributed internet data, sparking ethical and copyright concerns. Notable instances include lawsuits from artists and authors whose work was used without permission or compensation. Their contributions have significantly enhanced these models' capabilities, yet they receive neither credit nor financial reward. This scenario offers little motivation for African contributors to participate. Proposals for alternative models, like those suggested by experts such as Jaron Lanier[15] or TeHiku Media[16], aim to foster new data economies, recognizing contributors' roles and rights. However, their widespread uptake remains uncertain. One solution could be developing distinct African models, either as standalone entities or as auxiliaries of existing base models, incorporating more contextual relevance and ensuring data attribution and compensation.

# Takeaways and Recommendations

**To realize a vision of AI tailored for Africa, several unique features must be considered:**

### 1. Communal Focus Over Individualism:

**Existing generative AI technologies often embody Western perspectives, given that they are conceived of and built in the Global North. There is an opportunity to rethink this and embrace communal decision-making and community-oriented values. For example, taking into account data ownership and governance modalities that draw from Africa's community-centric ethos, and which work to counteract and avoid extractive practices, are likely to be beneficial to communities globally. What would Africa-centric models, which balance the needs of communities and individuals look like? What type of innovations around data ecosystems could arise from taking a less extractive perspective? Some places to look for inspiration include Sabelo Mhlambi's work on Ubuntu as an Ethical and human rights AI framework[17]; and the African Union's charter on Human and People's rights[18] which emphasizes the individual and community aspect to rights and introduces much overlooked language on duties/responsibilities to balance assertions of rights. The methods and algorithms created in building models to reflect such concerns are likely to be globally relevant and beneficial.**

### 2. Diverse Data Representation:

**AI models must be trained on data reflective of Africa's vast cultural and linguistic diversity, overcoming the current bias towards Western contexts. This will ensure AI's effectiveness and relevance across the continent. More equitable data ecosystems need to be created to enable this.**

15 University of California Television (UCTV) (2023) Data dignity and the inversion of AI. https://www.youtube.com/watch?v=itpbLcaW5WI.
16 Hao, K., 2022. A new vision of artificial intelligence for the people. MIT Technology Review. https://pulitzercenter.org/stories/new-vision-artificial-intelligence-people
17 Mhlambi, S., 2020. From rationality to relationality: ubuntu as an ethical and human rights framework for artificial intelligence governance. Carr Center for Human Rights Policy Discussion Paper Series, 9, p.31. https://cyber.harvard.edu/story/2020-07/rationality-relationality-ubuntu-ethical-and-human-rights-framework-artificial
18 Organization of African Unity (1981). African Charter on Human and Peoples' Rights. Nairobi: OAU. Available at: https://au.int/sites/default/files/treaties/36390-treaty-0011_-_african_charter_on_human_and_peoples_rights_e.pdf (Accessed: [6/7/2024]).

**3. Empowerment in Economic Models:**
Whilst there is a tendency to lean towards automation when technologies embody the Western perspective, AI in Africa should support the substantial informal sector, emphasizing empowerment, entrepreneurship, and job creation rather than mere efficiency. Again, this approach to Generative AI is likely to become globally relevant.

**4. Bridging the Digital Divide:**
AI development must account for varying levels of technological infrastructure across Africa, ensuring accessibility and reducing the digital divide. For example, by focusing on building smaller, less computationally heavy models that can run on basic smartphones, feature phones or leverage USSD and SMS services.

**5. Sustainable and Societal Focus:**
AI in Africa should prioritize sustainable development and long-term community well-being, diverging from the Western emphasis on short-term commercial success.

**6. Incorporating Traditional Knowledge:**
Respecting and integrating Africa's rich traditional knowledge and practices is crucial, offering a more inclusive and contextually relevant AI approach.

These key features highlight the need for AI models that are not just technically efficient but also culturally sensitive, and socially responsible, reflecting Africa's needs and values. In addressing these challenges, Africa has the opportunity to lead the way globally by working towards using Generative AI to create a more sustainable, human-centric, and value-driven future or work.

**Respecting and integrating Africa's rich traditional knowledge and practices is crucial, offering a more inclusive and contextually relevant AI approach**

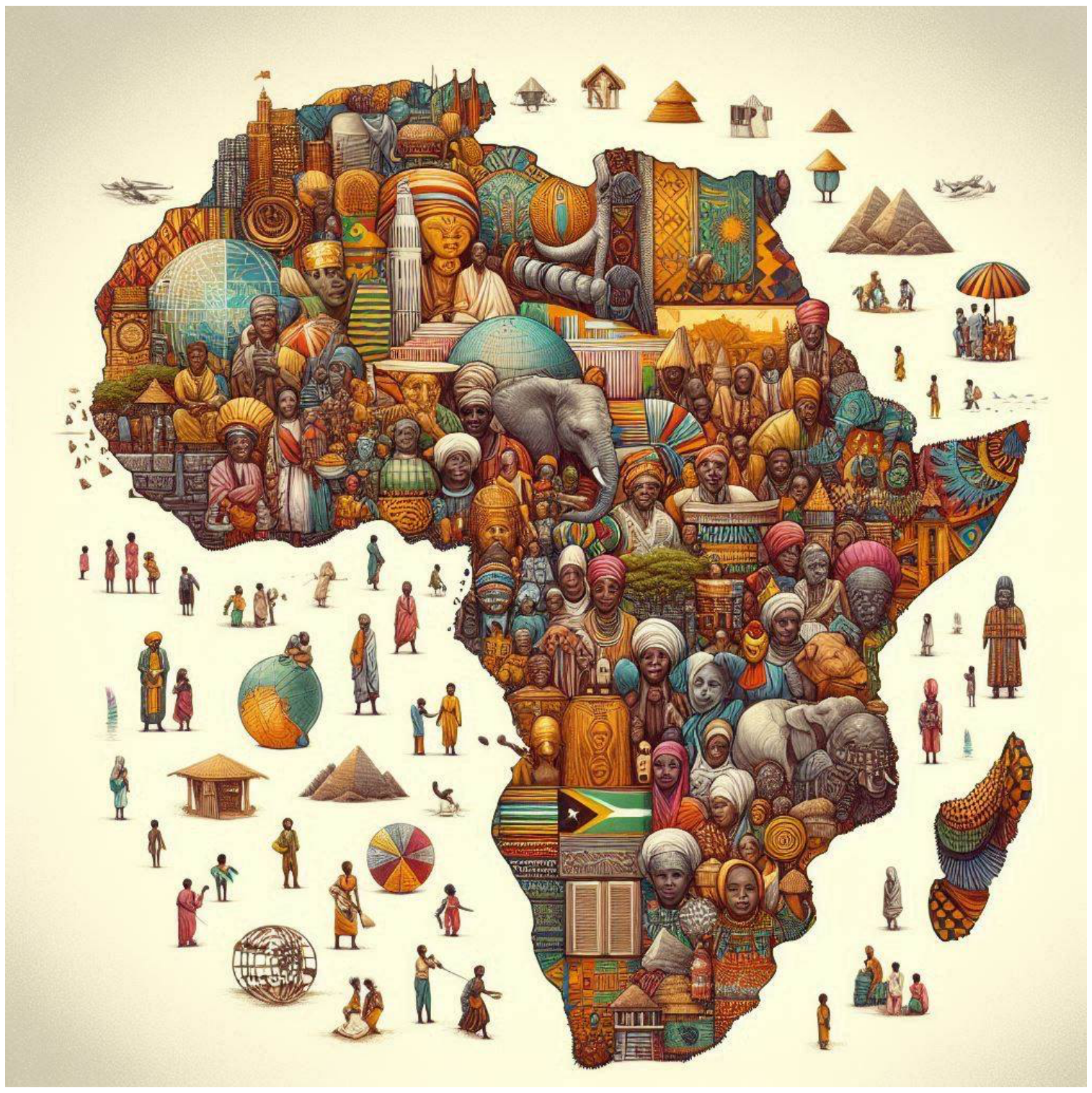

# Africa-Centric AI Tools and Platforms

## Introduction

Africa-centric AI refers to the design, development, validation and deployment of AI solutions with a strong focus on African contexts starting from problem definition and resource distribution. By leveraging local expertise and context, Africa-centric AI endeavors to foster innovation and collaboration, unlocking opportunities for prosperity across the continent. The emergence of Africa-centric AI tools and platforms addresses unique socio-economic and demographic opportunities and challenges by tailoring AI solutions to the continent's specific needs. These initiatives aim to improve access to essential services and drive economic growth.

Generative artificial Intelligence presents both opportunities and challenges in the context of Africa-centric AI tools and platforms. The opportunities lie in the potential for AI to enhance creativity, innovation, and problem-solving across various sectors in Africa. For instance, Generative AI can facilitate the development of locally relevant content, in the technical and scientific realms by enabling the creation of customised solutions for specific challenges faced by African societies, including in healthcare[1], agriculture[2], finance[3], disaster management and education. However, along with these opportunities, AI also brings challenges that need to be addressed. These challenges include technological barriers related to infrastructure and access to computing resources, energy requirements for training and deploying generative models, security concerns regarding data 1 Owoyemi, Ayomide, Joshua Owoyemi, Adenekan Osiyemi, and Andy Boyd. "Artificial intelligence for healthcare in Africa." Frontiers in Digital Health 2 (2020): 6.

2 Gwagwa, Arthur, Emre Kazim, Patti Kachidza, Airlie Hilliard, Kathleen Siminyu, Matthew Smith, and John Shawe-Taylor. "Road map for research on responsible Artificial Intelligence for Development (AI4D) in African countries: The case study of agriculture." Patterns 2, no. 12 (2021).

3 Mhlanga, David. "Industry 4.0 in finance: the impact of Artificial Iintelligence (AI) on digital financial inclusion." International Journal of Financial Studies 8, no. 3 (2020): 45.

**privacy and algorithmic bias, connectivity issues in remote areas, and the future implications of AI on employment and the workforce in Africa. Additionally, there are concerns about bias in AI algorithms, data scarcity, labour fairness, toxic language, and misinformation, concerns which are amplified in the African context. For example, in countries like Kenya, there have been several challenges with the treatment of workers involved in data collection for training AI models[4]. Many workers face exploitative working conditions and lack protection from psychological harm. These challenges, which may be more acutely felt on the African continent due to existing socio-economic disparities and infrastructural limitations, underscore the importance of addressing ethical and societal considerations. It is essential for stakeholders to collaboratively work towards mitigating these challenges while maximising the opportunities presented by Generative AI for the benefit of African communities and beyond.**

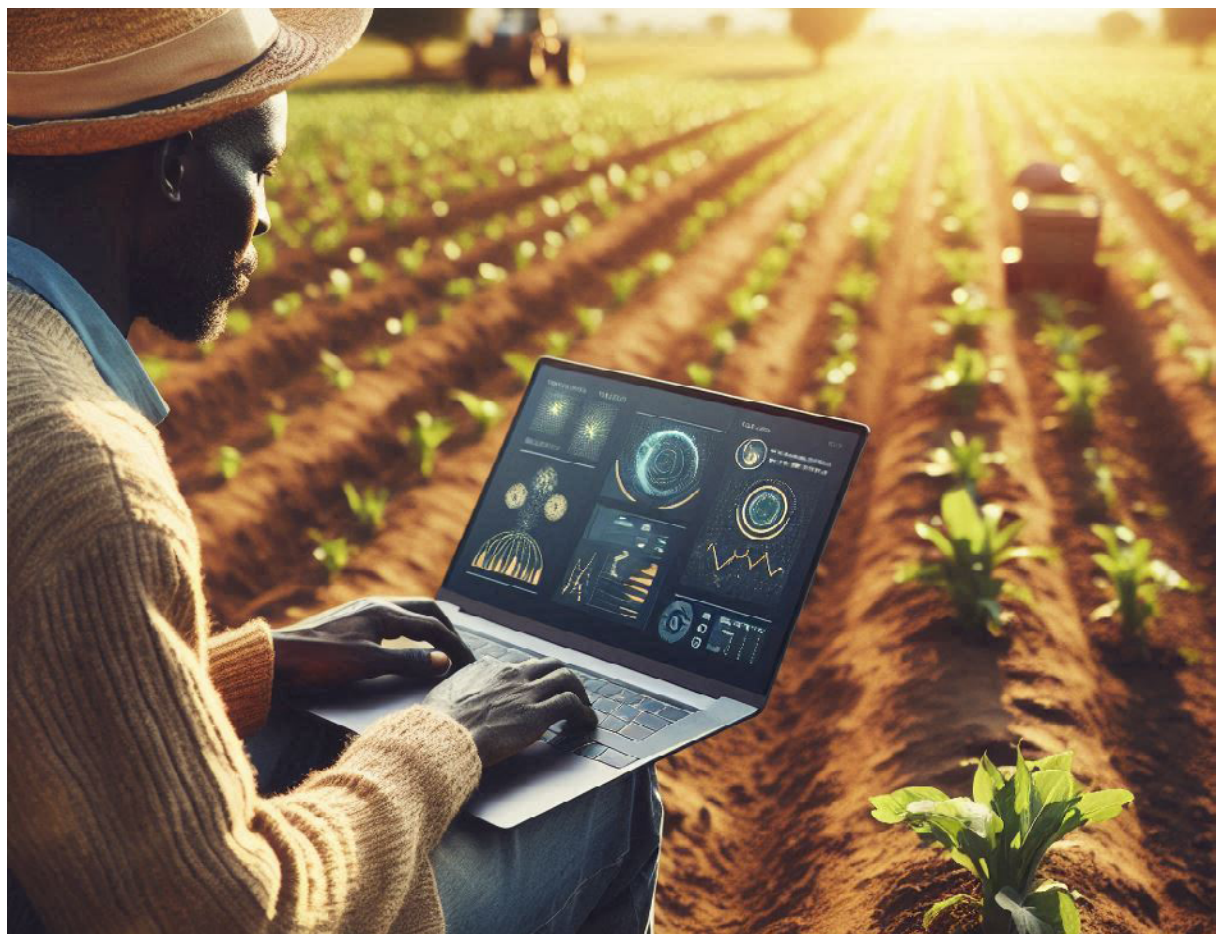

**Whilst some opportunities and challenges related to AI are globally common, such as data governance and ethical considerations, their solutions may not be. Furthermore, issues like connectivity, access to electricity, and language diversity pose unique challenges in the African context. This requires crafting solutions to take into account the unique context and needs of African communities. We need to ensure that the needs, context and voices of Africans are understood and considered in the development of these technologies. This includes building models that are fit for purpose on the continent; better data governance that allows for model and data sovereignty; and holding organizations to account for automated decision-making using AI models. All of the above will require African research, development and innovation that develops AI tools and platforms that are Africa-centric, as well as building just African AI ecosystems. A major example of this need is the lack of African language representation in Large Language models (LLMs)[5], which is connected to how data for such models is collected[6], as well as the lack of available resources for African language development in AI.**

## Generative AI in Africa: Prospects and Considerations

As digital transformation takes root across Africa, the emergence of AI technologies brings a spectrum of opportunities that can catalyze progress across various sectors. The growing performance of generative AI, particularly language models, could play a significant role in utilising the continent's rich tapestry of languages and cultures, a unique challenge that Generative AI can convert into an asset. For example, an AI system could provide personalised fertiliser recommendations for a farmer in the rural parts of Africa, tailored to her/his farm conditions and expected climate patterns, but also using the farmer's mother tongue. Similar examples could be drawn in the education sector, where a student anywhere in Africa could access a personalised educational resource, including a tutor, in the student's native language. By enabling real-time translation and content localisation, AI can ensure that linguistic diversity becomes a conduit for inclusion rather than division. This can revolutionise education and

4 Mhlanga, David. "Industry 4.0 in finance: the impact of artificial intelligence (ai) on digital financial inclusion." International Journal of Financial Studies 8, no. 3 (2020): 45.
5 Mhlanga, David. "Industry 4.0 in finance: the impact of artificial intelligence (ai) on digital financial inclusion." International Journal of Financial Studies 8, no. 3 (2020): 45.
6 Mhlanga, David. "Industry 4.0 in finance: the impact of artificial intelligence (ai) on digital financial inclusion." International Journal of Financial Studies 8, no. 3 (2020): 45.

information access, leading to increased literacy and a more informed populace.

Such advancements could improve living standards, foster economic growth, and bridge the digital divide. However, it's essential to strike the right balance and implement checks to maximize the benefits of the technology while mitigating potential adverse impacts.

In the realm of infrastructure and logistics, AI stands to introduce unprecedented efficiency. It could transform supply chain management, improve public service delivery, and improve governance by enhancing accountability and transparency. This is particularly important for Africa, where infrastructure can often be a bottleneck to economic activity. In the longer term, assuming sustained infrastructure development, one could imagine intelligent systems which can enhance service delivery by optimizing the distribution of essential services like electricity, water, and the Internet.

AI's role in social development is equally significant. In healthcare, it could enable better healthcare support and even predictive care models. In Agriculture, it could lead to smarter, more efficient farming practices. Education systems can benefit from personalised AI-driven learning approaches, potentially transforming the educational landscape and outcomes for students. The focus on research and development investment is a positive indication of the intent to build a dynamic AI sector within the continent. Such investment could catalyse innovation, creating high-value jobs, and establishing Africa as a competitive force in the global AI arena advancing beyond the current paradigm practice, where Africa is considered a cheaper source of alternative for data labelling labour. Moreover, AI's alignment with environmental sustainability is critical for Africa - particularly due to the severe impact climate change poses to the continent, an impact which is disproportionate to its emissions. Smart energy systems powered by AI could facilitate the transition to sustainable energy sources, promoting environmental resilience in the face of climate change. The fact that much of the energy in Africa comes from renewables, e.g., >90% for Kenya, positions the continent to be a leading alternative to host high computing centres for sustainable AI.

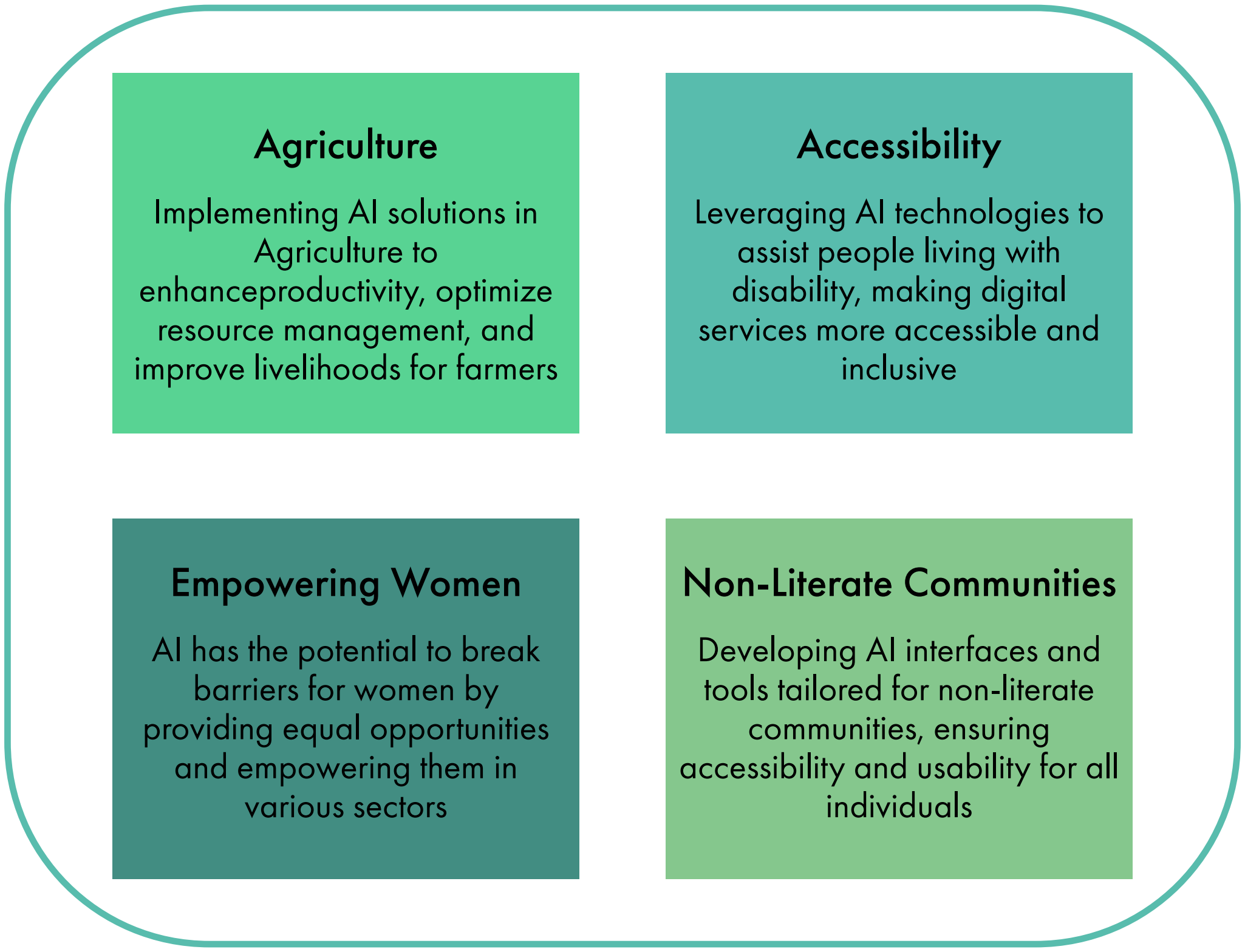

*Figure 1: Inclusivity in AI: A Priority*

While Africa boasts a rapidly growing interest in AI, particularly Generative AI, its ascent is accompanied by a suite of challenges that demand proactive management to avoid exacerbating existing societal gaps. It's crucial to remember that Africa is not a monolith. The continent encompasses 54 diverse countries at various stages of development. Literacy rates and language disparities can significantly limit AI accessibility for substantial portions of the population across these countries. User-friendly text-free or reduced text interfaces designed for low-literacy are essential to bridge this gap. Connectivity remains a hurdle, particularly in remote areas where internet access is limited, further hindering the potential benefits of AI innovations. Unaddressed, AI may exacerbate the digital divide, disproportionately impacting disadvantaged populations. This underscores the need for parallel investments in infrastructure to ensure comprehensive and equitable AI integration across the continent.

Furthermore, the potential negative societal impacts of AI, particularly Generative AI using synthesized content, cannot be understated. Issues like discrimination, fake news, misinformation, hate speech[7], and polarization present significant challenges, especially for African countries with fragile democracies and a history of instability. Mitigating these risks requires a multi-pronged approach that emphasizes responsible AI development, media literacy initiatives, and robust regulatory frameworks.

There is a danger of increasing marginalisation for vulnerable groups such as women, Persons with Disabilities, less literate people, people living in rural areas with limited or no connectivity, and smallholder farmers if AI tools are not tailored to their specific needs. For example, although language technology may increase accessibility to people who are not literate or have learning disabilities, potential beneficiaries need basic hardware, internet connection, and skills. Further, due to various social factors including deep-set patriarchy, women in Africa have poorer access to digital technologies than men[8]. Ensuring that AI is an enabler for these groups is crucial for fostering an inclusive technological ecosystem. The potential displacement of workers due to AI and automation is also a significant concern, especially in sectors susceptible to such shifts. This necessitates foresight in policymaking, focusing on education reform and vocational training to prepare the workforce for the changing job landscape. The impact of automation on job loss for African women, who occupy the majority of positions in low-skilled labour and repetitive tasks, needs attention[9].

**Ensuring that AI is an enabler for vulnerable groups is crucial for fostering an inclusive technological ecosystem**

The promising opportunities of AI could be achieved in Africa, only if all stakeholders are involved along the life cycles of AI: problem identification, impact analysis, user identification, AI solution design, development, evaluation and deployment, and temporal calibration. The stakeholders' group shall be extended to reflect both the understanding of the problem on the ground and the benefits of AI adoption but also to be aware of the concerns early in the process. This is essential in order to ensure the trustworthiness of AI solutions and hence increase the likelihood of these solutions being used by respective users, e.g., clinicians and policymakers. Data privacy and security emerge as paramount issues in the context of AI. As AI systems feed on large datasets, establishing robust data protection laws and ethical frameworks is essential for maintaining public trust and ensuring responsible AI use. Lastly, the distribution of investments in AI must be strategic to avoid reinforcing economic disparities. It is essential to balance these investments with other developmental priorities to ensure broad-based growth and equitable benefit distribution.

7 Weidinger, L. et al., (2021). Ethical and social risks of harm from Language Models: https://arxiv.org/pdf/2112.04359.pdf
8 Adams, R. (2022). AI in Africa: Key Concerns and Policy Considerations for the Future of the Continent: https://afripoli.org/ai-in-africa-key-concerns-and-policy-considerations-for-the-future-of-the-continent
9 Tadesse, G. A., Ogallo, W., Cintas, C., Speakman, S., Walcott-Bryant, A., & Wayua, C. (2024). Bridging the gap: leveraging data science to equip domain experts with the tools to address challenges in maternal, newborn, and child health. npj Women's Health, 2(1), 13.

A world that integrates generative AI can diverge onto vastly different paths depending on how the technology is developed, deployed, and governed. The outcome could range from dystopian to utopian, according to the design and deployment choices made, ethical considerations, and policy decisions. In the workshop, we envisaged both scenarios as a way of probing what Africa-Centric AI might look like. We present the two scenarios below.

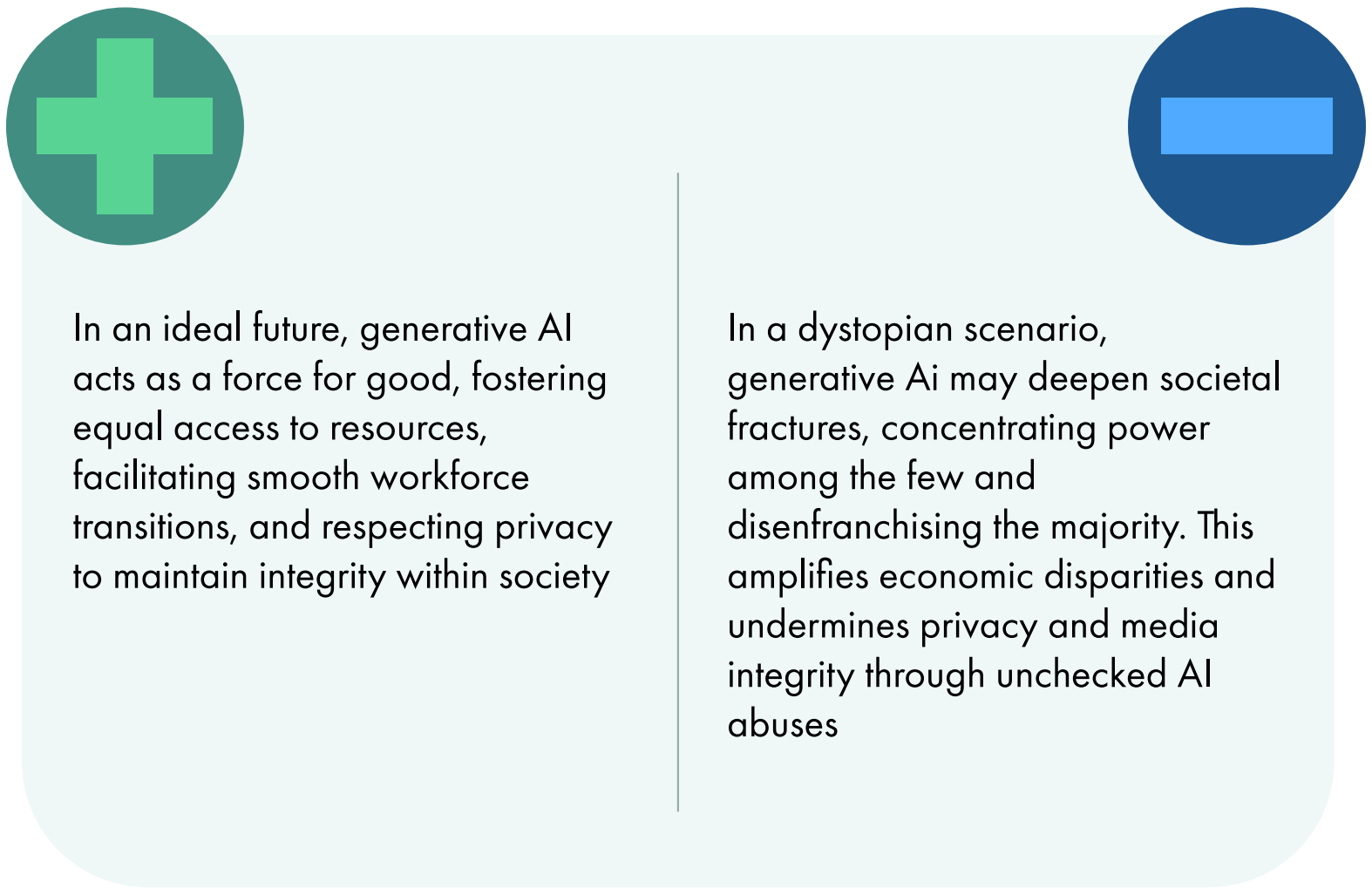

*Figure 2: Dystopian vs Utopian AI Analysis*

## A Dystopia with Generative AI

In a less ideal scenario, Generative AI could exacerbate existing societal fractures[10]. If left unchecked and unregulated, these AI systems could deepen the divide between the digital "haves" and "have-nots." In this world, AI accelerates the concentration of wealth and power in the hands of the few who control the technology, leaving the majority with only limited access to its benefits. The economic disparity widens as job displacement due to automation rises without adequate systems in place for re-skilling or social safety nets. Privacy erosion becomes rampant as AI technologies capable of generating and synthesizing vast amounts of personal data are misused for illegitimate surveillance by despotic governments, fraud, scams, or manipulative advertising. Such misuse would see an increase in the spread of misinformation and hate speech as Generative AI becomes a sophisticated tool for bad actors to create convincing fake content, undermining trust in media and institutions. In such a world, AI-driven decisions, lacking transparency and accountability, could lead to discriminatory practices, with algorithms reinforcing biases in everything from law enforcement to hiring practices. The unchecked environmental impact of powering extensive AI systems contributes to ecological degradation as energy demands soar without sustainable practices in place.

Generally, AI systems that are designed to be inclusive of global cultures, respect cultural differences, and perform effectively across various cultural contexts are crucial for their deployment both in Africa and beyond[11]. However, current practices in AI data collection[12], evaluation, design, and deployment often overlook the diversity of global cultures and fail to acknowledge the cultural values influenced by AI. Without addressing this gap, there's a risk of promoting AI systems that are not aligned with Africa's cultures thereby failing to get adoption and as a consequence failing to realise any positive societal impact[13].

10 Birhane, Abeba. "Algorithmic colonization of Africa." SCRIPTed 17 (2020): 389.
11 Birhane, Abeba, William Isaac, Vinodkumar Prabhakaran, Mark Diaz, Madeleine Clare Elish, Iason Gabriel, and Shakir Mohamed. "Power to the people? opportunities and challenges for participatory AI." Equity and Access in Algorithms, Mechanisms, and Optimization (2022): 1-8.
12 Okorie, Chijioke. "Copyright, Data Mining and Developing Models for South African Natural Language Processing." (2023).
13 Moorosi, Nyalleng, Raesetje Sefala, and Sasha Luccioni. "AI for Whom? Shedding Critical Light on AI for Social Good." In NeurIPS 2023 Computational Sustainability: Promises and Pitfalls from Theory to Deployment. 2023.

### A Utopia with Generative AI

Conversely, in a world where AI is harnessed for the greater good, technology can act as an equaliser. AI can democratise access to information and services, providing customised education and healthcare, regardless of geographical location or socioeconomic status. The workforce would transition smoothly from jobs vulnerable to automation to those that AI cannot, or should not, replicate - jobs requiring human creativity, empathy, and complex problem-solving.

In this optimistic scenario, Generative AI is developed with an ethical framework, respecting privacy and prioritising the security of not only personal data, but also communal data, developing and deploying communal data ownership models which preserve the interests and rights or indigenous and marginalized communities. Misinformation would be combated with AI tools that can verify facts and sources, helping to maintain the integrity of the information landscape. Algorithms in this world are transparent and regularly audited for bias, ensuring fairness in automated decisions. The deployment of AI is environmentally conscious, with AI systems optimised for energy efficiency and powered by renewable resources. Society benefits from AI's ability to enhance human capabilities, not replace them. Technology is leveraged to tackle global challenges such as climate change, poverty, and disease. It fosters a culture of collaboration and creativity, leading to breakthroughs in science, art, and governance.

### The Path Forward

The trajectory towards either of these futures is determined by the actions taken today. Ensuring a beneficial outcome with Generative AI involves proactive governance, inclusive design, investment in education, and a commitment to regulatory and ethical standards. The choice of which world we steer towards is a collective responsibility, requiring engagement from policymakers, technologists, and citizens alike. It is through thoughtful stewardship of AI technologies that we can aim for a future where such technologies serve to uplift and unite rather than divide and degrade.

# Takeaways and Recommendations

## What Do We Need To Do To Ensure A Positive Future of Work Using Generative AI?

Firstly, integrating AI into the workforce must be underpinned by a commitment to ethical development, transparency, and bias mitigation. This approach will ensure that AI systems are designed to enhance human capabilities without discrimination, fostering an inclusive environment where AI tools are accessible to all individuals. Moreover, the educational sphere must cultivate lifelong learners who possess not only technical skills but also essential human traits such as creativity, empathy, and ethical judgement. Ethical considerations are key to incorporating the Africa-related cultures and sources of knowledge into the AI framework.

Robust regulatory oversight is essential to govern the ethical deployment of AI, balancing innovation with safeguards against potential misuse that could harm the job market or infringe upon worker rights. Additionally, the potential disruption of certain job sectors by AI necessitates the strengthening of social safety nets to support individuals whose jobs are transformed by AI, ensuring a dignified transition. This journey towards a positive future with AI requires a collaborative effort involving governments, businesses, educational institutions, civil society, technical communities and academia to shape policies that promote a fair, equitable, and opportunity-filled job market. To this end, it is critical to engage in frequent and multi-stakeholder conversations (e.g., workshops, symposiums, and debates) at different administrative layers in the continent. This helps to bring diverse views together as we plan to develop impactful Africa-centric tools and platforms.

Encouraging entrepreneurship and innovation will be crucial in uncovering new industries and job roles, positioning AI as a catalyst for new forms of employment rather than a replacement for human work. Thus, more funding is necessary to shape the conversation and encourage ethical development and usage of AI solutions across different sectors.  Furthermore, global cooperation is imperative, necessitating the co-development of AI solutions, sharing of knowledge, alignment on ethical standards, and collaborative research to ensure widespread benefits and collectively addressing challenges related to AI in the workplace.

An ongoing conversation about the implications of AI for work, enriched by education on its capabilities and limitations, will help set realistic expectations and prepare society for the transitions ahead. Policymakers must anticipate the waves of change AI will bring and craft policies that encourage companies to invest in human-centric AI, guiding us toward a future where work is not just about productivity but about the potential for every individual to thrive alongside intelligent machines.

**Expanding the grassroots AI communities in Africa.** Compared to other regions where the AI ecosystem is shaped by big corporations or strong policies and regulation frameworks, Africa's AI ecosystem is dominated by grassroots movements, such as 'Deep Learning Indaba' and 'Data Science Africa'. Amplifying the work of and collaborating with these grassroots communities is critical to ensure the adoption of trustworthy AI in Africa. Local communities are already shown to be great platforms where entrepreneurs meet co-founders, communities of sub-specialties co-created, such as Masakhane (for Natural Language Processing), and students get mentorship and advising opportunities for further studies.

**Learning from practices outside Africa.** While Africa will benefit most from using tools and platforms that are Africa-centric, a critical look at external practices is also valuable to inform the Africa-centric development process. Incorporating benefits that align with Africa's contexts and discarding practices that do not. To this end, ensuring the trustworthiness and safety of existing AI solutions is of utmost importance. African researchers and practitioners should also aim for Africa-centric tools and developments that are tailored to the resources available e.g., learning from small data rather than the typical large datasets often employed for AI model training.

# Authors

This White Paper is a collective output from a wide number of people. The authors include:

Jacki O'Neill, Vukosi Marivate, Barbara Glover, Winnie Karanu, Girmaw Abebe Tadesse, Akua Gyekye, Anne Makena, Wesley Rosslyn-Smith, Matthew Grollnek, Charity Wayua, Rehema Baguma, Angel Maduke, Sarah Spencer, Daniel Kandie, Dennis Ndege Maari, Natasha Mutangana, Maxamed Axmed, Nyambura Kamau, Muhammad Adamu, Frank Swaniker, Brian Gatuguti, Jonathan Donner, Mark Graham, Janet Mumo, Caroline Mbindyo, Charlette N'Guessan, Irene Githinji, Lesego Makhafola, Sean Kruger, Olivia Etyang, Mulang Onando, Joe Sevilla, Nanjira Sambuli, Martin Mbaya, Paul Breloff, Dr. Gideon M. Anapey, Tebogo L. Mogaleemang, Tiyani Nghonyama, Muthoni Wanyoike, Bhekani Mbuli, Lawrence Nderu, Wambui Nyabero, Uzma Alam, Kayode Olaleye, Caroline Njenga, Abigail Sellen, David Kairo, Rutendo Chabikwa, Najeeb G. Abdulhamid, David Kairo, Ketry Kubasu, Chinasa T. Okolo, Eugenia Akpo, Tiyani Nghonyama, Joel Budu, Issa Karambal, Joseph Berkoh, William Wasswa, Muchai Njagwi, Rutendo Chabikwa, Rob Burnet, Loise Ochanda, Hanlie de Bod, Elizabeth Ankrah, Selemani Kinyunyu, Mutembei Kariuki, Angel Maduke, Kizito Kiyimba, Farida Eleshin, Lillian Secelela Madeje, Catherine Muraga, Ida Nganga, Judy Gichoya, Tabbz Maina, Samuel Maina, Muchai Mercy, Millicent Ochieng, Stephanie Nyairo

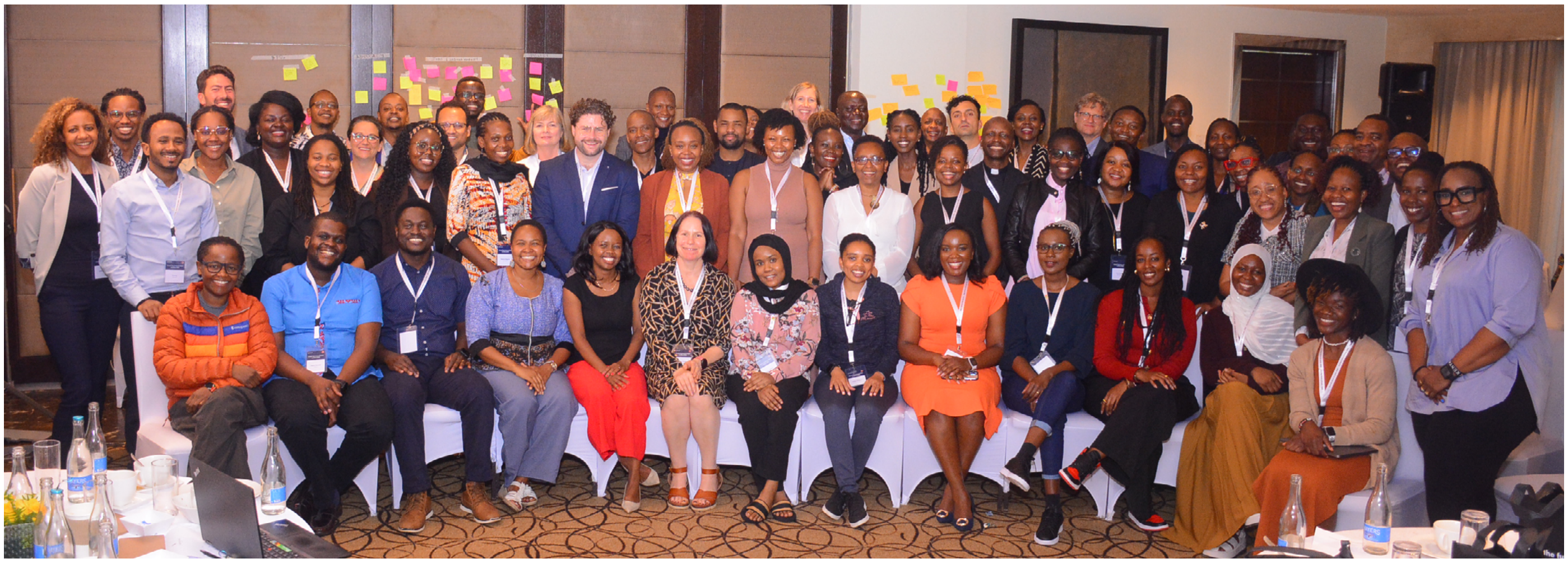

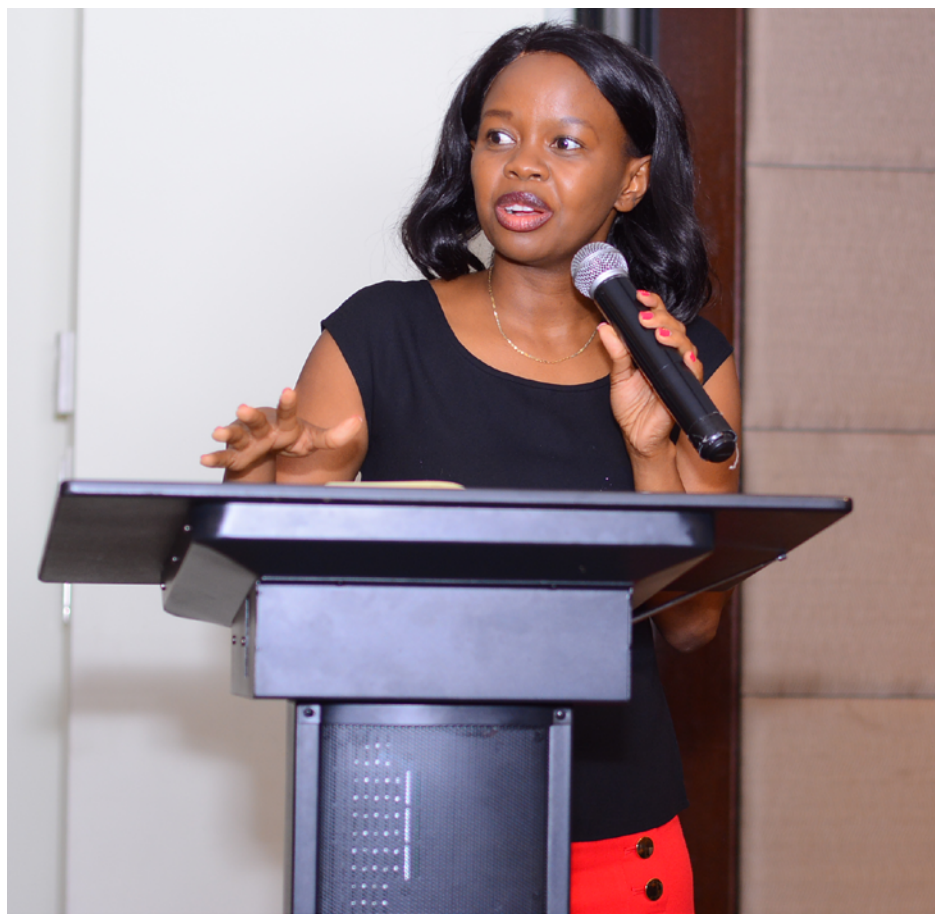

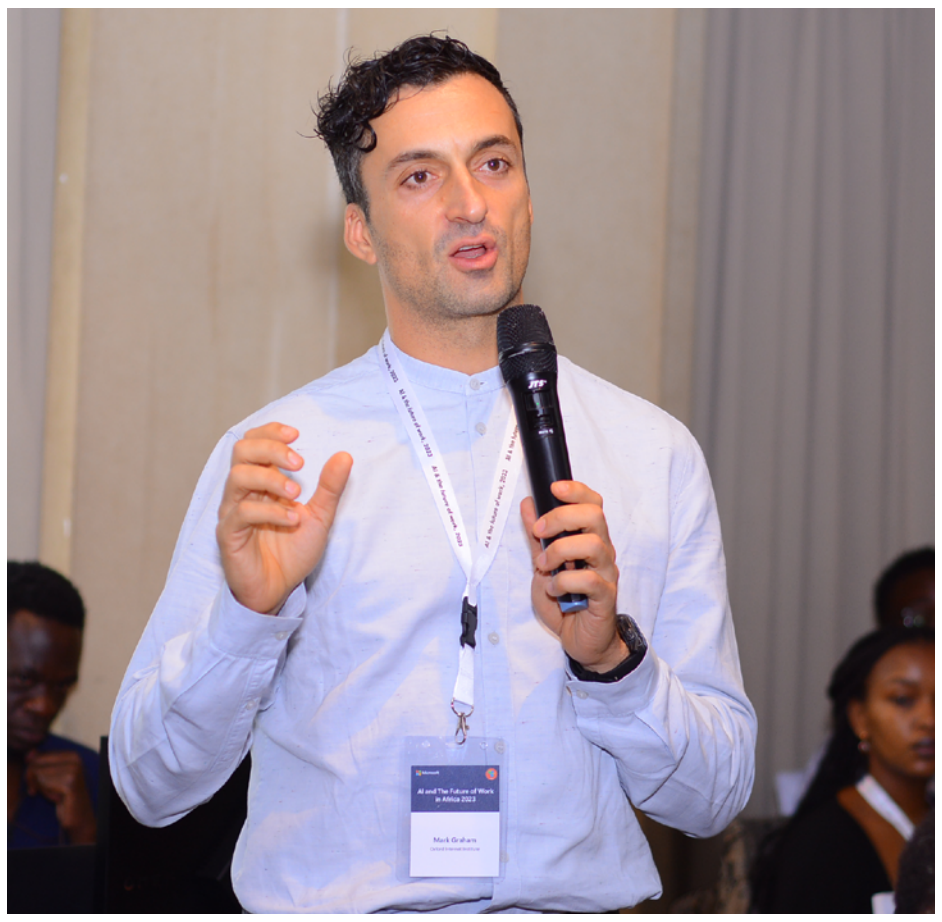

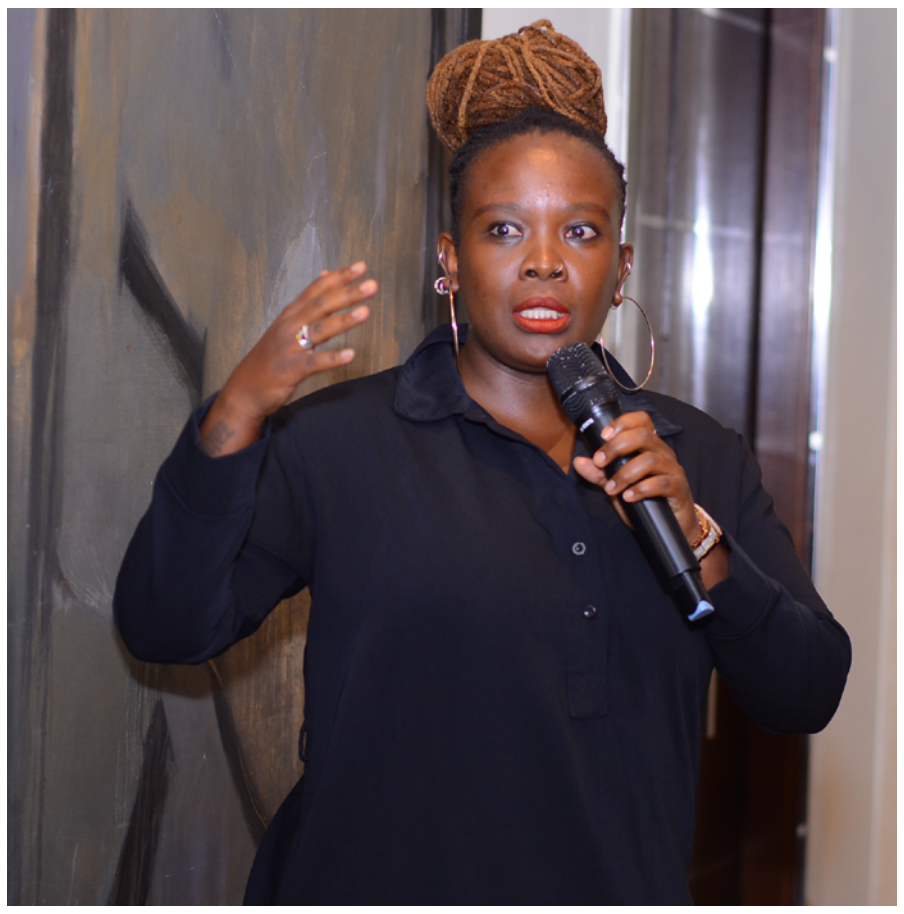

AI and the Future of Work in Africa

**White Paper**